\documentclass[10pt]{article}

\usepackage[hmargin=0.95in,vmargin=0.9in]{geometry}
\usepackage{amsmath,amssymb}
\usepackage[T1]{fontenc}
\usepackage{newtxtext,newtxmath}
\usepackage{booktabs}
\usepackage{graphicx}
\usepackage{array}
\usepackage{siunitx}
\usepackage{microtype}
\usepackage[font=footnotesize,skip=3pt]{caption}
\usepackage{titlesec}
\usepackage{enumitem}
\setlist{itemsep=2pt,topsep=3pt,parsep=0pt}
\titlespacing*{\section}{0pt}{1.6ex plus .6ex minus .3ex}{1.0ex plus .3ex}
\titlespacing*{\subsection}{0pt}{1.3ex plus .5ex minus .3ex}{0.8ex plus .2ex}
\titlespacing*{\paragraph}{0pt}{1.0ex plus .4ex minus .2ex}{1em}
\usepackage{xcolor}
\usepackage[hyphens]{url}
\usepackage[numbers,sort&compress]{natbib}
\usepackage[colorlinks=true,linkcolor=blue!60!black,citecolor=blue!60!black,urlcolor=blue!60!black]{hyperref}
\providecommand{\doi}[1]{\href{https://doi.org/#1}{doi:\nolinkurl{#1}}}

\graphicspath{{figures/}}
\newcommand{\sgn}{\operatorname{sgn}}
\newcommand{\sigmoid}{\operatorname{sigmoid}}
\newcommand{\VR}{\mathrm{VR}}
\newcommand{\AR}{\mathrm{AR}}

\begin{document}

\title{Short-horizon mean reversion in cryptocurrency markets:\\
a matched cross-market measurement}

\author{Nadav A. Kitron\\[2pt]
\small Independent researcher\\
\small \texttt{nadav.k@eshqol.com}
\and
Jonathan M. Wengrowicz\\[2pt]
\small Independent researcher\\
\small \texttt{yonatan.wengrowicz@gmail.com}}

\date{August 2026}
\maketitle

\begin{abstract}
At 15-minute horizons, directional mean reversion is far stronger and
more pervasive in cryptocurrency markets than in US equities: scored
under one matched, strictly out-of-sample protocol, 90\% of 183
Binance pairs carry significant directional reversal against 2.7\% of
187 US stocks and ETFs, in every focal coin-year since 2021. The
signal lives in signs, not magnitudes: lag-one return autocorrelation
is near zero on the major coins, yet simply betting against the
previous candle captures most of the effect. US-listed funds whose net
asset value is a crypto or metal price inherit their underlying's
reversal, including its absence, from their first months of trading;
stocks merely correlated with the same underlyings inherit nothing:
descriptive evidence that a wrapper's tape reads like the process it
wraps, not the venue it prints on. On the originating tape, the
reversal concentrates after moves driven by aggressive taker flow and
grows with flow intensity, while the order-book depth a move consumes
conditions nothing: a conditioning consistent with compensated
liquidity provision, not a test that selects it. The gross edge peaks
near 1.3\,bp per trade against a 5\,bp round-trip cost: large enough
to detect, too small to clear benchmark spot capture costs. The
contrast survives an artifact battery, an exact permutation null, and
a frozen six-month holdout, with a class-mean AUC gap of $+0.031$ as
designed and $+0.011$ (95\% CI $[+0.008,+0.014]$) under the most
conservative accounting, clear of zero either way.
\end{abstract}



\section{Introduction}\label{sec:intro}

Market efficiency is a matter of degree \citep{fama1970,grossman1980,
lim2011}: different markets absorb information at different speeds, and
the residual predictability left behind is a measurable signature of
their microstructure. This paper measures that residual at the shortest
horizons, at population scale, across today's two most different market
structures. Scored under one matched, strictly out-of-sample
walk-forward, 183 cryptocurrency pairs carry a pervasive 15-minute mean-reverting sign
predictability; in 187 US stocks and ETFs the residual reversal is too
weak to rank direction out of sample. The crypto signal is small (2--3
points of AUC), decays within hours, is stable across five years, and
survives every artifact channel we can test, a six-month
forward holdout, an exact permutation null, and the removal of the
common crypto market factor (Section~\ref{sec:fact}).

Two supporting analyses bound what the fact means. First, US-listed
funds whose net asset value \emph{is} a crypto or metal price reproduce their underlying's session-matched reading, nulls included,
while instruments merely correlated with the same underlyings carry nothing;
transmission is present from a fund's first months, through a
futures-based NAV, at $2\times$ leverage, and inside the sampling minute
(Section~\ref{sec:location}). This is descriptive evidence, not
identification: at one-minute correlation 0.982 a wrapper's bars nearly
\emph{are} its underlying's, so we read the panel as showing that
efficiency statistics computed on a wrapper's tape describe the wrapped
process, not the listing venue. Second, on the originating tape the
reversal concentrates after moves that aggressive taker flow pushed and
grows with flow intensity, while the net order-book depth a move
consumes conditions nothing: the signature of liquidity provision
compensating overreaction in a book that re-forms within the bar.
Because ordinary aggressor flow chooses to trade, no conditioning of
this kind can separate compensated provision from an informational
account, and we state the mechanism evidence at exactly that strength.
The gross edge peaks near
1.3\,bp per trade against a 5\,bp cheapest round-trip cost: big enough
to measure, too small to capture at benchmark spot costs, the residual
a limits-to-arbitrage account predicts
\citep{grossman1980,shleifer1997} (Section~\ref{sec:mech}).

The evidential hierarchy is stated once and held to. The population fact
rests on a test fixed in design before the wide-universe run and is the
firmest claim; the wrapper panel is an observational pattern on eight
underlying families, strong in direction and without formal
identification; the mechanism evidence constrains accounts without
selecting one: the flow conditioning is consistent with compensated
liquidity provision but cannot exclude informational alternatives
(\S\ref{sec:generator}). The measurement instrument is deliberately minimal: a three-parameter constrained distributed-lag logit used as a low-variance
device, not a model of markets. Nothing rests on it: an unconstrained
logit and parameter-free sign statistics reproduce the contrast, and its
derivation and baselines are confined to Appendix~\ref{app:model}.

\paragraph{Relation to the literature}
The paper sits in the microstructure literature on short-horizon reversal
and liquidity provision. In equities, reversal at daily and intraday
horizons has long been read as compensation for absorbing order-flow
imbalances \citep{lehmann1990,jegadeesh1990,grossmanmiller1988,
nagel2012}: strongest after high-volume moves \citep{campbell1993},
forecastable from signed order imbalance \citep{chordia2002}, tied to
illiquidity in the cross-section \citep{avramov2006}, estimated
structurally as inventory-driven price pressure \citep{hendershott2014},
and competed down over time by professional intermediaries
\citep{chordia2005,chordia2011}. We measure where that competition has
and has not done its work across today's market structures. The ETF
literature has used the creation--redemption link mainly in the
opposite direction: how arbitrage transmits non-fundamental shocks
\emph{into} underlyings \citep{bendavid2018}, how wrapper prices
deviate from NAV \citep{petajisto2017,madhavan2016}, how ETF activity
affects the underlying's efficiency \citep{glosten2021}, and where
price discovery sits among linked instruments \citep{hasbrouck1995};
we take the NAV constraint as given and ask whether a wrapper carries
its underlying's \emph{conditional} short-horizon statistic. On the
crypto side, prior work documents short-horizon inefficiency and
reversal with unconditional scaling statistics
\citep{urquhart2016,bariviera2017,kristoufek2018,alvarez2018,drozdz2018,
watorek2021}, prices a weekly short-term reversal factor
\citep{liu2022}, and describes a fragmented, retail-heavy structure
\citep{makarov2020,kogan2024}. Closest to our reading, a cross-sectional
literature earns daily-rebalanced reversal returns across coins and
interprets them as compensation for liquidity provision, with the
premium concentrated in small, illiquid, low-activity pairs
\citep{zaremba2021,bianchi2022,farag2025}. None of these ingredients is
new on its own, and the cross-sectional studies differ from ours on
every margin that matters here: they rank coins against one another at
daily and weekly horizons and earn most of their premium in the thin
tail, where liquidity-provision returns are amplified. We measure a
within-pair, sign-level conditional statistic at 15-minute horizons,
undiminished in the most liquid pairs where that cross-sectional
premium is weakest, against a matched US-equity control scored under the
identical strictly out-of-sample protocol, with a frozen forward
holdout and permutation-calibrated inference. The estimator's constrained-kernel form has a lineage in the
econophysics literature on spin models of markets
\citep{bouchaud2013,campajola2019,campajola2021}, set out in
Appendix~\ref{app:model}; none of the empirical results depend on it.

\section{Data, measurement device, and protocol}\label{sec:protocol}

\paragraph{Data}
The primary universe is 15-minute candles over 2025-01-01 to 2026-02-11:
the 183 highest-volume Binance USDT spot pairs and 187 liquid US
stocks/ETFs (24-hour bars; regular-trading-hours restrictions where
stated). Every dataset below is scored by the same models, walk-forward
geometry and bootstrap as the primary universe, and none is reused to
select any other's design. The supporting datasets, each introduced
where used, are: a focal panel of fifteen instruments across six asset
classes at 5\,m--4\,h; multi-year 15-minute histories back to 2021-01;
independent-venue refetches (Coinbase, OKX, Bybit) and quote-midprice
bars (Dukascopy); the wrapper panel of Section~\ref{sec:location} with
its launch-era and one-minute legs; Binance taker-flow, order-book and
perpetual series for Section~\ref{sec:mech}; and the
frozen universe refetched over 2026-02--2026-08 as the post-sample
holdout. Each bar carries the
intra-bar return $r_t=(\text{close}_t-\text{open}_t)/\text{open}_t$ and
label $\text{up}_t=\mathbf{1}[r_t>0]$; bars with $r_t=0$ (\emph{flat
bars}) take label 0 under this convention, which is an artifact channel
Section~\ref{sec:fact} treats explicitly. The convention pushes the
median up-rate to 0.458 in crypto and 0.420 in stocks (one stock bar
in six is labelled ``down'' by tie), so every headline statistic is
the base-rate-invariant AUC \citep{hanley1982} and the flat-bar
channel is carried as a separate effect size throughout.
Construction, cleaning, and survivorship details are in
Appendix~\ref{app:data}.

\paragraph{Measurement device}
The primary model is an autoregressive logit on the previous $N=12$
soft-clipped returns $s_{t-k}=\tanh(\lambda r_{t-k})$, with lag weights
tied to a decaying kernel:
\begin{equation}
P(\text{up}_t)=\sigmoid\!\Big(C+A\sum_{k=1}^{N}k^{-\alpha}\,s_{t-k}\Big),
\label{eq:model}
\end{equation}
three free parameters $(C,A,\alpha)$ regardless of $N$. We call this the
\emph{constrained logit} throughout. The amplitude $A$ carries the
economics: $A<0$ is mean reversion, $A>0$ momentum, $\alpha$ the memory
decay. The kernel tie is what makes it a low-variance instrument:
tying the weights to $A\,k^{-\alpha}$ forces the whole vector onto a
two-dimensional monotone manifold, so effective degrees of freedom do
not grow with $N$ (L$_1$/L$_2$ penalties, by contrast, treat every lag
alike). Appendix~\ref{app:model} derives the restriction,
notes its correspondence with a kinetic Ising chain, shows the data
identify the kernel's decay but not its functional form, and reports
the baselines the tie beats on out-of-sample log-loss (one-lag, free,
norm-penalized, tuned machine-learning). An
unconstrained free $\AR(12)$ logit is fit alongside everywhere, and a
result is \emph{two-model significant} only when both clear the test,
so ``predictable'' always means two different estimators agree.
A third, parameter-free statistic accompanies them: the kernel-weighted
sign correlation
\begin{equation}
R=\operatorname{corr}\!\Big(\sigma_t,\ \textstyle\sum_{k=1}^{12}
k^{-\alpha}\sigma_{t-k}\Big),\qquad \sigma_t=\sgn(r_t),\quad
\alpha\equiv 1,
\label{eq:R}
\end{equation}
with $\alpha$ fixed a priori and nothing fitted anywhere: the
model-free counterpart used wherever a fitted result needs
corroboration.

\begin{center}
\fbox{\begin{minipage}{0.94\linewidth}
\small
\textbf{The primary test.}
\emph{Universe:} 183 Binance USDT spot pairs (highest 24-hour quote
volume at selection, excluding stablecoin-quote pairs and leveraged
tokens) and 187 liquid US stocks/ETFs; exact symbol lists frozen in the
replication package.
\emph{Data:} 15-minute candles, 2025-01-01 to 2026-02-11; label
$\mathbf{1}[r_t>0]$ on intra-bar open-to-close returns.
\emph{Models:} the kernel-constrained logit of Eq.~\eqref{eq:model}
($N=12$, $\lambda=150$) and the unconstrained free $\AR(12)$ logit, both
by MLE with selection on a chronological validation tail.
\emph{Walk-forward:} train 5{,}760 / test 960 candles, step equal to the
test block; non-overlapping out-of-sample blocks concatenated and scored
once.
\emph{Statistic and inference:} out-of-sample AUC; one-sided
moving-block-bootstrap $p$-value for AUC $>0.5$ (block 384 candles,
$B=300$ resamples) per model, so the attainable $p$-floor is
$1/(B{+}1)=0.0033$; per-asset conjunction
$p_{\cap}=\max(p_{\text{constr}},p_{\text{free}})$; Benjamini--Hochberg
FDR at $q=0.05$ jointly across all 370 assets.
\emph{Population claim:} class-mean AUC gap under a joint moving-block
bootstrap on the shared time grid ($B=1{,}000$, raised to $5{,}000$ in
the sensitivity runs of Appendix~\ref{app:protocol}).
\emph{Exclusions:} assets with fewer than
train${}+{}$test${}+{}2{,}000$ candles, and assets with fewer than
2{,}000 out-of-sample candles; none other.
\end{minipage}}
\end{center}

\paragraph{Protocol and inference}
Everything is scored strictly out of sample under the boxed protocol
\citep{kunsch1989,politis1994,benjamini1995,harvey2016};
\emph{session-matched} regular-trading-hours (RTH) series, with
24-hour tapes and listed instruments alike restricted to the identical
New York 09:30--16:00 slot grid, use a 60/10-trading-day walk-forward.
The population statement is a \emph{joint} moving-block bootstrap on
the shared time grid, which preserves cross-sectional dependence
(Section~\ref{sec:fact} states the estimand exactly). The primary test
(universe, span, models, walk-forward geometry, statistic, FDR step) was
fixed in design before the wide universe was run; the wrapper panel and
the mechanism experiments were designed afterwards, on instruments
outside its universe, and are reported as what they are. A refit-inclusive
bootstrap, negative controls, and all protocol details are in
Appendix~\ref{app:protocol}.

\section{The population fact: reversal concentrates in crypto relative
to US equities}\label{sec:fact}

\begin{figure}[htbp]
\centering
\includegraphics[width=0.72\linewidth]{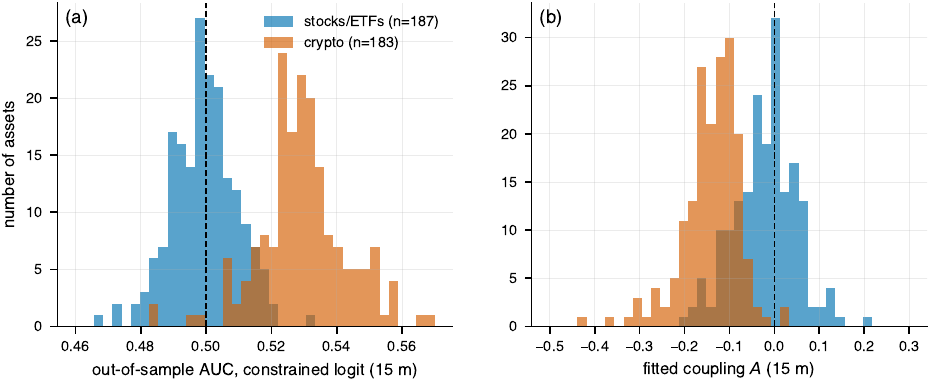}
\caption{The primary test: 15-minute out-of-sample skill across 370
assets (183 Binance crypto pairs, 187 US stocks/ETFs), identical
walk-forward and inference. (a)~AUC by class; the separation shown is
the all-bars scoring, and narrows under the artifact-robust rescorings
of \S\ref{sec:fact}. (b)~Fitted coupling $A$: mean-reverting ($A<0$)
for crypto, centered on zero for stocks.}
\label{fig:widedist}
\end{figure}

Figure~\ref{fig:widedist} is the population-level fact. Crypto AUC
averages 0.531 with 98\% of pairs above 0.5; stocks average 0.499 and
split evenly around it, and after joint FDR control 90\% of crypto pairs
are significant against 2.7\% of stocks (164 of 183 against 5 of 187).
The joint moving-block bootstrap puts the
class-mean gap at $+0.031$ (95\% CI $[+0.027,+0.035]$, stable across
bootstrap block lengths), an exact permutation null that preserves the
full dependence structure places the observed gap 4.4 null SDs out,
beyond all 340 admissible label shifts (Appendix~\ref{app:perm}), and
the fitted coupling is mean-reverting in 98\% of crypto pairs against a
zero-centered stock distribution. The few crypto pairs
without a signal are dominated by pegged stablecoins and brand-new
microcaps; the few US-listed instruments clearing the raw test are led
by the metal ETFs, which are listed claims on round-the-clock
underlyings, a point Section~\ref{sec:location} makes systematic.

Because crypto pairs co-move so strongly, the estimand deserves an
exact statement: the \emph{difference in cross-sectional mean
out-of-sample AUC between the two selected universes}, each asset's
AUC computed once on its concatenated out-of-sample series over this
window. The ``90\% of 183 pairs'' tally describes how broadly the
signal covers that cross-section; it is not 183 independent
replications, since with mean pairwise correlation 0.40 the effective
number of independent crypto observations is roughly 2.5 (stocks:
correlation 0.23, ${\approx}4.2$), which is why all inference runs
through the joint bootstrap. The defensible reading is ``this liquid-crypto
cross-section carried a small but pervasive and persistent reversal
signal throughout the window,'' not ``183 markets independently
confirmed one.'' Small is the right word: 3.1 points of AUC before
artifact adjustment, 2.2 after, 52.3\% unthresholded accuracy on BTC.
What earns the report is breadth and persistence, not size.

\paragraph{An ex-ante universe, reported in full}
The one researcher degree of freedom the fixed design could not remove
is the universe itself: the 183 pairs were ranked by volume measured
\emph{after} the sample ended, so the as-fixed test runs on a
survivor-tilted membership. We therefore report, with equal prominence,
an \textbf{ex-ante specification} that removes the late-listing channel:
the 133 of the 183 pairs already trading in the first sample month,
re-ranked by that month's volume. The candidate pool is still the
post-sample top 183, so this bounds the late-listing tilt rather than
eliminating the volume ranking itself. Every headline statistic
is unchanged or stronger on it. The class-mean gap is $+0.035$ (95\% CI
$[+0.030,+0.038]$) under the constrained logit and $+0.030$
$[+0.026,+0.033]$ under the free logit; after joint FDR control, 97\%
of the 133 pairs are significant against the same 5 of 187 stocks
(2.7\%) as the primary specification (the equity leg is identical); the
coupling is negative in 98\% (mean $-0.15$); and across ex-ante volume
quintiles the AUC is flat, with a slightly \emph{negative} volume--AUC
rank correlation ($-0.16$). Late-listed survivors dilute rather than
drive the effect, and predictability does not live in thin pairs. Both
universes still condition on surviving to the download date (delisted
pairs are unrecoverable from the public API); within what public data
allow, the contrast does not depend on the part of the selection rule
we can vary.

\paragraph{A six-month post-sample holdout}
Every modelling choice in this paper was frozen against data through
2026-02-11; roughly six months later we refetched the identical frozen
universe over 2026-02-12 to 2026-08-08 and reran the pipeline on
outcome data that did not exist when any choice was fixed. Three sample-size thresholds were
relaxed for the shorter span, and only those: minimum bars 8{,}720 to
6{,}960, minimum out-of-sample count 2{,}000 to 480, and the joint
bootstrap's within-replicate minimum from 1{,}000 to 300; the last is not a bar cut and bears on the
interval below. The holdout FDR family is
the 225 assets scored. No model, geometry, statistic or block length
changed. The effect is attenuated but clearly present
(Table~\ref{tab:holdout}): the joint dependence-preserving gap is
$+0.020$ (95\% CI $[+0.010,+0.028]$) against $+0.031$ in-sample, on
roughly a third of the original sample length, with 60\% of scored
crypto pairs still FDR-significant. The parameter-free sign statistic $R$ of
Eq.~\eqref{eq:R} is nearly unchanged on the crypto side
($-0.034$ against $-0.035$ in-sample, negative in 96\%), while the
stock side moves from $-0.000$ to $-0.006$ (66\% negative), so the
model-free class gap
attenuates by a fifth against the fitted gap's third: the attenuation
is mostly, not wholly, in the fitted model's out-of-period calibration.
The ex-ante subset reads the same (mean AUC 0.521).

Three qualifications. Universe membership is not clean of the holdout
window (the frozen list was ranked on volume at a date inside it), so
the holdout tests the overfitting objection, not the selection
objection, which the ex-ante specification addresses. The model legs
cover liquidity subsets while the model-free statistic covers
everything and is the leg to read for the full stock universe. And the
attenuation decomposes: rescoring the \emph{primary} window on exactly
the holdout-scored assets, 23\% of it is the universe change and 77\%
decay on a fixed universe (\texttt{holdout\_did}). One six-month
window cannot settle the adaptive-markets question, but the
overfitting objection is tested directly and fails.

\begin{table}[htbp]
\centering
\footnotesize
\setlength{\tabcolsep}{6pt}
\begin{tabular}{@{}lcc@{}}
\toprule
& Primary (2025-01--2026-02) & Holdout (2026-02--2026-08) \\
\midrule
Class-mean gap, constrained & $+0.031$ $[+0.027,+0.035]$ & $+0.020$ $[+0.010,+0.028]$ \\
Class-mean gap, free logit & $+0.026$ $[+0.023,+0.029]$ & $+0.015$ $[+0.008,+0.021]$ \\
Model-free sign-gap ($R$, crypto $-$ stocks) & $-0.035$ & $-0.028$ \\
Crypto mean AUC & 0.531 & 0.522 \\
Stock mean AUC & 0.499 & 0.501 \\
Crypto pairs FDR-significant & 90\% & 60\% \\
Crypto coupling $A<0$ & 98\% & 96\% (mean $-0.21$) \\
\bottomrule
\end{tabular}
\caption{The frozen pipeline on post-sample data. Gaps carry 95\% joint
moving-block-bootstrap CIs; the holdout model legs are the liquidity
subsets described in the text, and the model-free row covers the full
universes.}
\label{tab:holdout}
\end{table}

The focal panel (Table~\ref{tab:main}) places the two ends of the
population contrast in a wider cross-section of market structures.
Under the two-model criterion the tally is crypto 3/3, spot FX 2/3,
US-listed 1/8: the three coins clear the no-skill line easily, the
dealer-intermediated European FX pairs show a faint copy, and
exchange-listed equities, indices, and Treasuries show nothing. That
ordering runs along intermediation of the primary market rather than
along an equity/crypto dichotomy; the wrapper panel of
Section~\ref{sec:location} traces it further. (The one US-listed
exception, GLD, is itself a wrapper on an OTC underlying: on
regular-trading-hours bars it reads null, like its underlying, the
concordance Section~\ref{sec:location} builds on.)
Focal-panel detail: Appendix~\ref{app:focal}.

\begin{table}[htbp]
\centering
\footnotesize
\setlength{\tabcolsep}{4pt}
\begin{tabular}{@{}llrrcccrr@{}}
\toprule
Instrument & Class & $n_{\text{OOS}}$ & AUC & 95\% CI & Constr.\ $p$ & Free $p$ & $A$ & $\rho_1$ \\
\midrule
ETH & crypto & 33{,}312 & 0.538 & [0.532, 0.544] & $\mathbf{<10^{-3}}$ & $\mathbf{<10^{-3}}$ & $-0.31$ & $+0.005$ \\
XRP & crypto & 33{,}312 & 0.536 & [0.530, 0.542] & $\mathbf{<10^{-3}}$ & $\mathbf{<10^{-3}}$ & $-0.27$ & $-0.023$ \\
BTC & crypto & 33{,}312 & 0.533 & [0.527, 0.539] & $\mathbf{<10^{-3}}$ & $\mathbf{<10^{-3}}$ & $-0.33$ & $-0.003$ \\
GLD & commodity & 11{,}642 & 0.520 & [0.506, 0.531] & \textbf{0.005} & $\mathbf{<10^{-3}}$ & $-0.10$ & $+0.024$ \\
GBPUSD & fx & 21{,}928 & 0.516 & [0.510, 0.523] & $\mathbf{<10^{-3}}$ & \textbf{0.002} & $-0.74$ & $-0.023$ \\
EURUSD & fx & 21{,}930 & 0.514 & [0.507, 0.523] & $\mathbf{<10^{-3}}$ & \textbf{0.014} & $-0.71$ & $-0.021$ \\
NVDA & stock & 11{,}996 & 0.512 & [0.501, 0.524] & 0.015 & 0.258 & $-0.13$ & $+0.013$ \\
USDJPY & fx & 21{,}934 & 0.509 & [0.502, 0.516] & 0.008 & 0.239 & $-0.13$ & $-0.010$ \\
AAPL & stock & 11{,}940 & 0.505 & [0.495, 0.514] & 0.176 & 0.055 & $-0.13$ & $-0.008$ \\
QQQ & index & 11{,}994 & 0.501 & [0.493, 0.510] & 0.353 & 0.116 & $-0.07$ & $-0.006$ \\
TSLA & stock & 11{,}996 & 0.501 & [0.490, 0.508] & 0.506 & 0.211 & $-0.05$ & $+0.007$ \\
IWM & index & 11{,}975 & 0.497 & [0.485, 0.507] & 0.744 & 0.138 & $-0.05$ & $-0.033$ \\
TLT & bond & 11{,}917 & 0.492 & [0.483, 0.502] & 0.927 & 0.643 & $-0.07$ & $+0.008$ \\
SPX & index & 12{,}006 & 0.492 & [0.482, 0.500] & 0.972 & 0.641 & $-0.05$ & $-0.030$ \\
\bottomrule
\end{tabular}
\caption{The focal panel at 15\,m: out-of-sample AUC (constrained
logit), 95\% CI, one-sided $p$ for AUC$>0.5$ under both models, fitted
coupling $A$ (fixed $\lambda$; standardized values in
Appendix~\ref{app:robust}), and lag-1 return autocorrelation. Bold:
$p<0.05$ under both models. SOL lacks matched 15-minute data here; the
independent-venue refetch fills the gap (AUC 0.529, both models
$p<10^{-3}$).}
\label{tab:main}
\end{table}

\begin{figure}[htbp]
\centering
\includegraphics[width=0.5\linewidth]{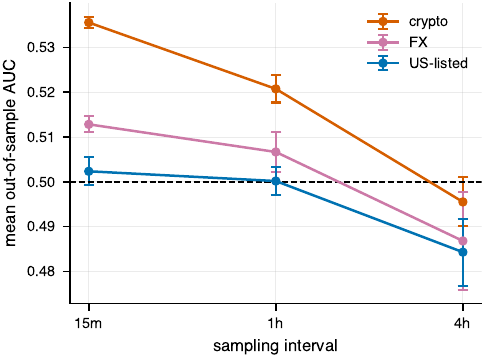}
\caption{Mean out-of-sample AUC vs.\ sampling interval ($\pm$SE across
assets, focal panel). The signal is strongest at the highest frequency
and decays to no-skill within hours; the class ordering (crypto $>$ FX
$>$ US-listed) holds at every horizon. Per-instrument 15-minute values:
Table~\ref{tab:main}.}
\label{fig:focal}
\end{figure}

\begin{figure}[htbp]
\centering
\includegraphics[width=0.82\linewidth]{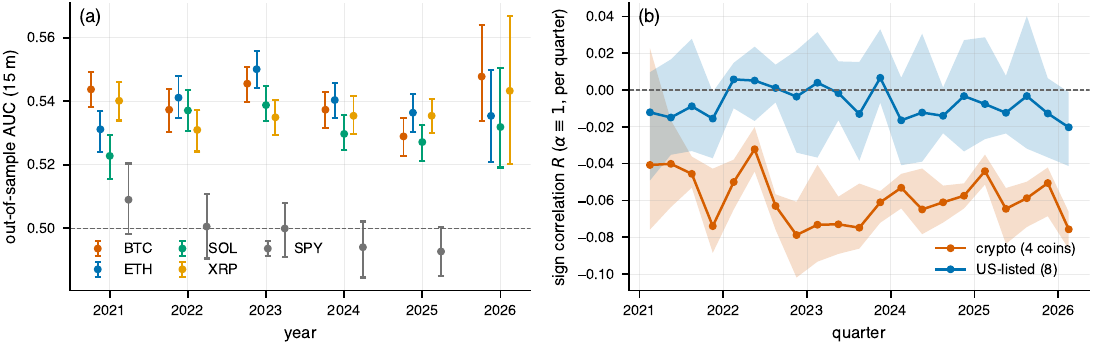}
\caption{Five years of the same effect, read two ways. (a)~Per-year
out-of-sample AUC of the constrained model, four focal coins,
2021--2026 (95\% bootstrap CIs; 2026 is a partial year), with SPY
scored identically as the no-skill reference. (b)~The model-free sign
correlation $R$ of Eq.~\eqref{eq:R} per calendar quarter (class mean;
shading spans the min--max across assets): the crypto mean is negative
in all 21 quarters, the US-listed mean stays near zero.}
\label{fig:stability}
\end{figure}

\paragraph{Finite memory, standing effect}
Across sampling
intervals (Figure~\ref{fig:focal}), the signal decays monotonically
and is gone by four hours: the profile of a microstructure effect with
finite memory rather than of slow-moving mispricing. At population scale the
hourly rerun keeps 58\% of the 169 crypto pairs meeting the hourly
bar-count cut significant after FDR versus 0\% of stocks. Across
calendar time it is a standing feature (Figure~\ref{fig:stability}):
every focal coin-year since 2021 is two-model significant, the sign
correlation is negative in every quarter, and SPY scored identically
never leaves the no-skill band. The contrast is therefore not an artifact of
the 14-month primary window or of any single volatility regime. The mild
downward drift in BTC's per-year AUC is the one feature consistent with
slow adaptive erosion \citep{kristoufek2018,khuntia2018,lim2011}, and
Section~\ref{sec:discussion} states the corresponding falsification
test.

\begin{figure}[htbp]
\centering
\includegraphics[width=0.78\linewidth]{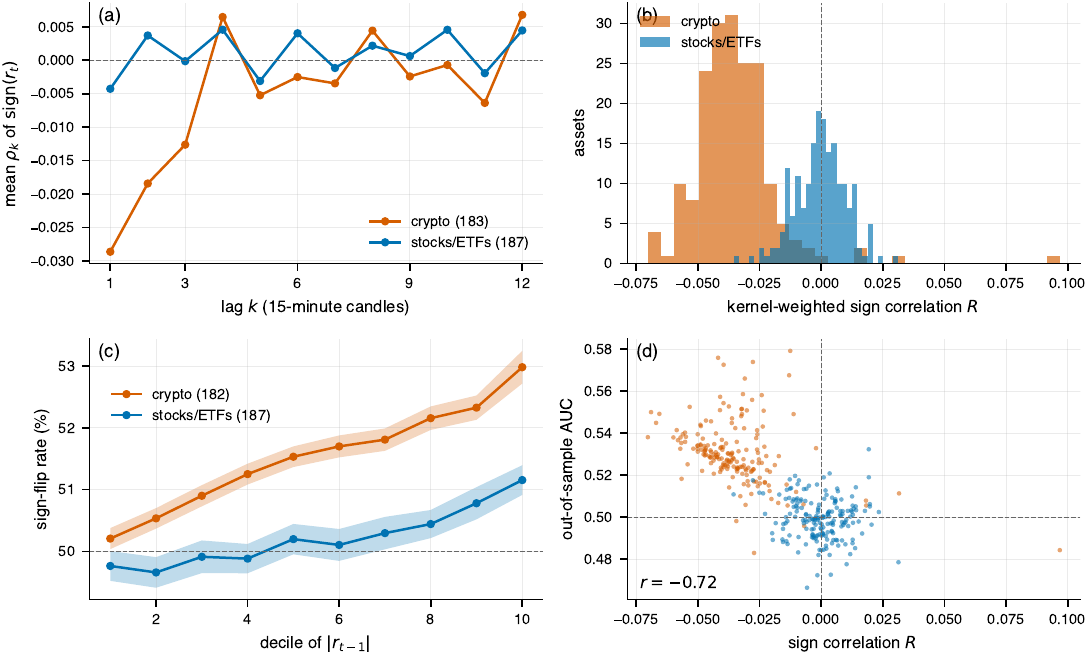}
\caption{The same contrast with no fitted model. (a)~Class-mean
autocorrelation of $\sgn(r_t)$ by lag $k$. (b)~Histograms of the
kernel-weighted sign correlation $R$ (Eq.~\eqref{eq:R}, $\alpha\equiv1$
fixed a priori), one count per asset. (c)~Class-mean sign-flip rate by
decile of the preceding move's size $|r_{t-1}|$, with shaded dispersion
bands (one pegged pair with degenerate deciles is excluded from (c); 182
pairs). (d)~$R$ against the fitted model's out-of-sample AUC,
one point per asset over all 370 ($r=-0.72$). Panels (a), (b) and (d)
use all 183 crypto pairs.}
\label{fig:signlag}
\end{figure}

\paragraph{The contrast is a property of the data, not the estimator}
The lag-1 \emph{return} autocorrelation is ${\approx}0$ on the focal
coins, which makes the signal look invisible to the standard first
check. The resolution (Figure~\ref{fig:signlag}) is that short-horizon
crypto reversal is a \emph{sign} phenomenon that strengthens with the
size of the preceding move: the sign-flip rate rises monotonically
across all ten deciles of $|r_{t-1}|$, from a base above $50\%$ in every
one ($50.2\to53.0\%$), and because $\rho_1$ weights bar
pairs by $r_{t-1}r_t$, the top decile's contribution nearly cancels the
middle deciles'. Stocks share the gradient at half the slope and from a
lower base ($49.8\to51.2\%$, non-monotone, at or below $50\%$ through
the fourth decile), so the size-dependence is not crypto-exclusive, only
markedly larger there. Measured on signs over many lags the structure is not
weak at all: the wide-universe class-mean lag-1 autocorrelations are
$-0.046$ (returns) and $-0.029$ (signs) against $-0.009$ and $-0.004$
for stocks, the multi-lag sign autocorrelation is significant in 97\%
of crypto pairs versus 13\% of stocks, and the parameter-free sign
correlation $R$ of Eq.~\eqref{eq:R} reproduces the class gap at
$-0.035$ (95\% CI $[-0.041,-0.029]$) under the identical joint
bootstrap, negative in 98\% of crypto pairs against 50\% of stocks,
with nothing fitted anywhere. The focal coins carry the same reading
one instrument at a time (sign $\rho_1$ $-0.038$ to $-0.056$,
Ljung--Box $p<10^{-15}$, against return $\rho_1$ never above $0.022$
in magnitude).
Hurst exponents from detrended fluctuation analysis (DFA)
corroborate at population level, with crypto antipersistent and a
class-mean $H$ gap of $-0.049$ (95\% CI $[-0.084,-0.012]$) under the
same joint bootstrap (Appendix~\ref{app:robust}).

The same discipline applies to the kernel's own length
(Table~\ref{tab:nbaseline}): a one-lag
special case, the probabilistic form of betting against the previous
candle's sign, already reaches a crypto class-mean AUC of 0.527
against the twelve-lag kernel's 0.531. The population fact is therefore
predominantly a lag-one reversal with a short memory tail;
Appendix~\ref{app:nbaseline} quantifies the kernel's small, systematic
increment and its saturation by $N\approx3$.

\begin{table}[htbp]
\centering
\footnotesize
\setlength{\tabcolsep}{5pt}
\begin{tabular}{@{}p{0.27\linewidth}p{0.37\linewidth}p{0.28\linewidth}@{}}
\toprule
Artifact channel & Test & Result \\
\midrule
Trading-session mismatch & all 187 stocks on RTH-only bars &
gap intact (stocks 0.498) \\
Flat-bar labels & rescore on non-flat bars only &
gap $+0.022$ $[+0.019,+0.025]$ \\
One-bar bounce / staleness & one-bar gap between features and label &
71\% of excess survives \\
Thin-pair staleness & AUC vs.\ volume cross-section &
volume slope $\approx0$; flat quintiles \\
Single-venue idiosyncrasy & refetch from Coinbase, OKX, Bybit &
12/12 cells within $\pm0.002$ \\
Bid--ask bounce & quote-midprice bars (bid/ask average) &
unchanged (BTC 0.529, ETH 0.537) \\
Return convention & close-to-close, VWAP variants &
identical to third decimal \\
Quarter-hour clock effects & bars rebuilt at ${+}1/{+}2/{+}5/{+}7$\,min
offsets (focal coins, 1\,s data) & $R$ strongly negative at every
offset \\
Pipeline manufacture & shuffled labels, all 370 assets &
0/370 significant after FDR \\
Magnitude structure & sign-randomized surrogates, $|r|$ path fixed &
0/14 significant; crypto mean 0.496 \\
Regime dependence & per-year walk-forward 2021--2026 &
24/24 coin-years significant \\
Sample-tuned choices & frozen pipeline on post-sample
2026-02--2026-08 data & gap $+0.020$ $[+0.010,+0.028]$ \\
Estimator dependence & free logit; sign statistics; DFA &
gap $+0.026$; $-0.035$; $-0.049$ $[-0.084,-0.012]$ \\
Kernel length & one-lag ($N=1$) special case &
gap $+0.029$; $+0.004$ crypto gain to $N{=}12$ \\
Under-powered equity null & per-stock power; joint bound on class mean &
median power 0.98; mean $\le0.5007$ \\
Both channels at once & non-flat \emph{and} one-bar gap, jointly &
gap $+0.011$ $[+0.008,+0.014]$ \\
Inference machinery & exact permutation null, day-aligned label shift &
0/340 shifts reach the gap; $4.4$ null SD \\
Intraday-variance detectability & AUC within hour-of-day strata &
gap \emph{widens} to $+0.034$ \\
One common factor & equal-weight crypto factor projected out &
81\% of $R$ retained; 98\% still negative \\
\bottomrule
\end{tabular}
\caption{The artifact ledger. Each row is a mechanical or statistical
channel that could manufacture the crypto--equity contrast, the test
that isolates it, and the outcome. Full details, including the
phase-randomized surrogate control and its anti-conservativeness
diagnosis, are in Appendices~\ref{app:robust}
and~\ref{app:negcontrols}.}
\label{tab:ledger}
\end{table}

\paragraph{The fact survives every artifact channel}
Table~\ref{tab:ledger} compresses the robustness battery; three rows
carry the most weight: the full-universe RTH restriction (stocks stay
at 0.498, 24/7 crypto unchanged), the non-flat rescoring that removes
the one channel rewarding a reversal model with free ``correct'' calls
(crypto 0.531 to 0.520, stocks unmoved), and the one-bar
feature/label gap that removes boundary effects such as bid--ask
bounce \citep{roll1984} (71\% of the excess survives, 81\% of pairs
significant). Details: Appendix~\ref{app:robust}.

The paper carries two effect sizes, and they stay separate throughout.
The \textbf{primary analysis}, whose design was fixed ex ante
(\S\ref{sec:protocol}), is the all-bars gap of $+0.031$ (95\% CI
$[+0.027,+0.035]$). The non-flat rescoring is our \textbf{preferred
artifact-robust effect-size estimate}: crypto 0.5200 against stocks
0.4985, a class-mean gap of $+0.022$ (95\% CI $[+0.019,+0.025]$) under
the same joint bootstrap, and $+0.017$ $[+0.015,+0.019]$ under the free
logit. The latter is reached by a decision made after seeing the data,
so it never substitutes for the primary figure; its interval is well
clear of zero.

\paragraph{Both artifact channels at once}
One-at-a-time robustness overstates what survives, so we also report the
conjunction: the walk-forward refit with a one-bar feature/label gap,
scored with flat bars dropped, under the identical joint bootstrap
(\texttt{joint\_artifact}).
Both adjustments together leave crypto at 0.5093 against stocks 0.4980:
a class-mean gap of $\mathbf{+0.011}$ (95\% CI $[+0.008,+0.014]$). That
is 36\% of the headline gap, below the 47\% independent channels would
leave, so the two channels overlap: part of what the gap removes is structure
the flat-bar convention was manufacturing. The paper's effect
size is therefore a range, every point of it clear of zero: $+0.031$ as
fixed ex ante, $+0.022$ on either adjustment alone, $+0.011$ with both.

\paragraph{What the interval does and does not carry}
The headline interval is fit-conditional: blocks are resampled with the
fitted models held fixed, so it prices sampling variation but not
estimation uncertainty. Diagnostics that refit both models inside every
replicate cut the gap by a third to a half, to roughly $+0.017$ to
$+0.022$, and about double the interval width
(Appendix~\ref{app:protocol}). The permutation null quoted above needs
no interval at all, so the finding's significance does not rest on the
fit-conditional interval; $[+0.027,+0.035]$ is the precision of a
conditional estimate, not the paper's full uncertainty.

\paragraph{The equity null is powered}
Absence of evidence licenses evidence of absence only with a power
calculation, so we state one. Using each stock's own
moving-block-bootstrap AUC dispersion (the identical block geometry as
the per-asset test), the one-sided 5\% test would detect a true effect
of the crypto class-mean size ($+0.031$) with median power 0.98 across
the 187 stocks; power exceeds 0.8 in 94\% of them, with only the
shortest series falling to ${\approx}0.64$. The class-level statement is
sharper: the joint moving-block bootstrap puts a one-sided 95\% upper
bound of \textbf{0.5007} on the stock class-mean AUC, so the design does
not merely fail to find an equity-class effect; it bounds any common one
at less than a fortieth of the crypto class mean's excess
(Appendix~\ref{app:power}).

\paragraph{Two readings the design must exclude, and does}
First, \emph{detectability}. AUC ranks bars by a score that is a fixed
function of raw returns, so a steep intraday variance profile can bury a
real dependence: intraday heteroskedasticity is $17.6\times$ for listed
instruments against $5.0\times$ for crypto, and SPX, whose lag-1 return autocorrelation of $-0.030$ is ten times
BTC's, reads AUC 0.492 yet lifts to 0.515 when its variance is homogenized
(Appendix~\ref{app:negcontrols}). Could the whole
class gap be a detectability gap? No: recomputing each asset's AUC
within hour-of-day buckets, which removes the cross-regime pooling
exactly and needs no refit, \emph{widens} the gap from $+0.0313$ to
$+0.0335$ (\texttt{factor\_variance}). The check holds the fitted score
fixed, and the equity null remains a statement about a common effect
measured this way, not proof that US equities carry no linear reversal
at all; the
crypto side was never a candidate, since BTC's $\rho_1$ of $-0.003$
supports only 0.501 through the linear channel against 0.533 observed.

Second, \emph{one factor wearing 183 names}. With
$N_{\text{eff}}\approx2.5$ the cross-section could be a single
mean-reverting market factor. Projecting an equal-weight crypto factor
out of every pair (betas estimated on the first half of the span only)
and recomputing the parameter-free $R$ of Eq.~\eqref{eq:R} on the
residuals: the class mean moves from $-0.0358$ to $-0.0290$, retaining
81\%, and stays negative in 98\% of pairs (\texttt{factor\_variance}).
Four fifths of the sign reversal is idiosyncratic. The residuals are
not independent either, but the one-factor reading is ruled out.

\section{The signature in listed wrappers}\label{sec:location}

\begin{figure}[htbp]
\centering
\includegraphics[width=0.7\linewidth]{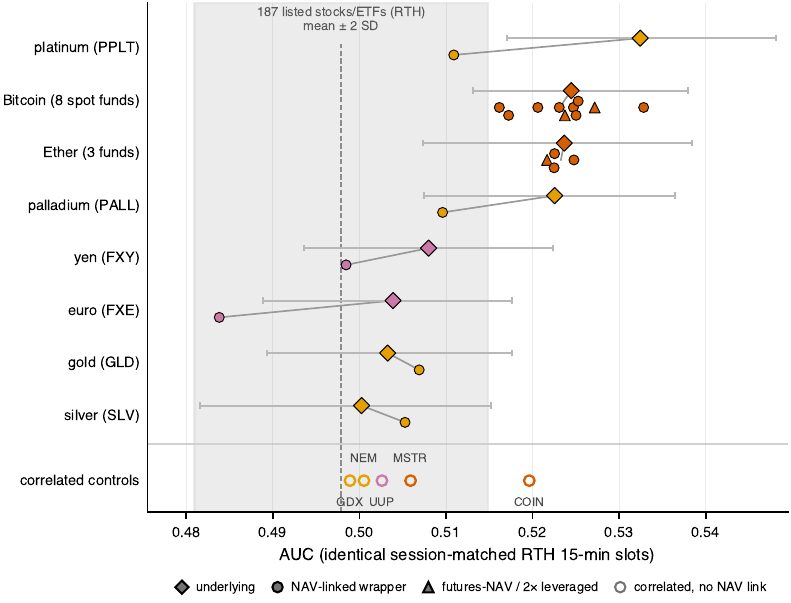}
\caption{The wrapper panel on session-matched RTH bars. Diamonds:
underlyings with 95\% bootstrap CIs; filled circles: their NAV-linked
wrappers (triangles: futures-NAV and leveraged funds); grey connectors
span the family gap; open circles: correlated controls; shaded band:
187-stock RTH reference (mean $\pm2$ SD). Live underlyings (Bitcoin,
Ether) are matched by their funds right of the band; null underlyings
(gold, silver, FX) sit in-band with theirs; the thin metal wrappers fall
back toward the band. Values: Table~\ref{tab:family},
Appendix~\ref{app:panel}.}
\label{fig:location}
\end{figure}

Does the reversal belong to the asset's price process or to the venue it
trades on? A cross-market contrast cannot say, since it varies both at
once. Listed wrappers vary them separately: eight underlyings (four spot
metals on OTC dealer networks, Bitcoin and Ether on 24/7 exchanges,
EURUSD and USDJPY on 24/5 spot FX) each have NAV-linked US-listed funds (seventeen spot-custody wrappers
plus the futures-based BITO and the $2\times$ leveraged BITU and ETHU),
all trading with designated market
makers, regular hours, and a consolidated tape. Five listed instruments
\emph{correlated with} but not priced off an underlying (GDX, NEM, COIN,
MSTR, UUP) separate ``NAV-linked'' from ``same risk factor.'' Every leg,
including the 24/7 and OTC underlyings, is scored on the identical New
York 09:30--16:00 15-minute slots under the identical walk-forward,
models, and bootstrap. Figure~\ref{fig:location} shows the panel;
Table~\ref{tab:family} aggregates it to the family level, where
observations are independent.

One limit governs the whole section, stated before the results: IBIT
and Bitcoin move together at $r=0.982$ per minute
(\S\ref{sec:leadlag}), so a liquid wrapper's 15-minute bars nearly
\emph{are} its underlying's, and inheritance on the live cells is
close to mechanical, under arbitrage the maintained hypothesis rather
than a discovery. What the panel adds is descriptive but useful: the
correlated controls carry nothing, the nulls are concordant, thin
tapes attenuate, and one corollary has teeth: efficiency statistics
computed on a wrapper's tape describe the wrapped process, not the
listing venue (Section~\ref{sec:discussion}).

\begin{table}[htbp]
\centering
\footnotesize
\setlength{\tabcolsep}{4pt}
\begin{tabular}{@{}llcccc@{}}
\toprule
Underlying family & Wrappers & Underl.\ AUC & Wrapper mean &
Family gap & SD units vs.\ stocks \\
\midrule
Bitcoin & 8 spot funds & \textbf{0.524} & \textbf{0.523} & $-0.001$ &
$+3.0$ \\
Ether & 3 funds & 0.524 & 0.523 & $-0.000$ & $+3.0$ \\
Gold & GLD & 0.503 & 0.507 & $+0.004$ & $+1.1$ \\
Silver & SLV & 0.500 & 0.505 & $+0.005$ & $+0.9$ \\
Platinum & PPLT (thin) & \textbf{0.532} & 0.511 & $-0.022$ & $+1.5$ \\
Palladium & PALL (thin) & \textbf{0.523} & 0.510 & $-0.013$ & $+1.4$ \\
Euro & FXE (thin) & 0.504 & 0.484 & $-0.020$ & $-1.7$ \\
Yen & FXY (thin) & 0.508 & 0.499 & $-0.010$ & $+0.1$ \\
\midrule
correlated controls & GDX, NEM, COIN, MSTR, UUP &
\multicolumn{4}{c}{0/5 significant; mean AUC 0.506} \\
\bottomrule
\end{tabular}
\caption{The panel at the family level. Funds on one underlying share
one price process and window, so they are within-family replications,
not separate tests; family gaps are computed at full precision. The last
column standardizes the wrapper mean against the 187-stock RTH
cross-section (mean 0.498, SD 0.0085), a descriptive scale rather than a
test statistic; the control COIN sits at $+2.6$ on it. Per-cell
inference: Table~\ref{tab:panel}. Bold: two-model significant at 5\%;
``thin'': $>15\%$ flat bars.}
\label{tab:family}
\end{table}

\subsection{Panel results}\label{sec:panelresults}

Where the underlying carries a signal on the session-matched slots, its
NAV-linked funds carry it too: the Bitcoin and Ether families match
their underlyings to a family gap of $0.001$ or less, three
cross-sectional SDs above the 187-stock reference. Where the underlying reads a null on those slots (gold and silver, whose
marginal 24-hour readings do not survive the session restriction), its
funds read the null. Instruments carrying the same risk without the NAV link carry
nothing: 0 of 5 controls, including COIN and MSTR, whose Bitcoin betas
dwarf any metal ETF's tracking error. The exposure itself is well
documented: listed Bitcoin-treasury firms co-move with BTC at an average
beta of $0.62$, a dozen of them, MSTR included, above one
\citep{aufiero2025}; what the controls lack is not the risk factor but
the NAV link. Thin tapes attenuate: PPLT and PALL read $0.013$--$0.022$
below their metals (Table~\ref{tab:family}; neither cell individually
significant), consistent with stale prints offsetting inherited
reversal, though without intraday NAV the panel cannot separate stale
sampling from partial non-transmission. The
omitted thinner Bitcoin and Ether funds could only widen the measured
family gaps, so the panel's composition flatters no conclusion.

Formal inference is weaker than the pattern, in three specific ways.
First, the eight Bitcoin funds share one process and one window: eight
within-family replications (SD 0.005, range 0.516--0.533), not eight
confirmations. Applying the wide study's own FDR discipline within the
33-leg panel family leaves 6 of 8 spot-Bitcoin funds, with Bitcoin
itself and the futures-NAV cell failing the corrected bar
(\texttt{panel\_fdr}). Second, regressing the
family wrapper mean on its underlying gives an inheritance slope of
$0.69$ (SE $0.32$): it rejects neither full inheritance ($p=0.37$) nor
\emph{zero} inheritance ($p=0.076$), so eight families cannot separate
the hypotheses the panel was built around, and we retire the permutation
$p$-values we previously reported. Third, COIN, kept on the control side because it fails free-logit
corroboration ($p=0.36$; constrained $p=0.001$), sits at $+2.6$ on the panel's own descriptive scale,
above every wrapper family except Bitcoin and Ether. Read strictly, the
panel separates live-underlying NAV-linked wrappers from everything
else, not NAV-linkage from correlation as such. Ether's cells are
likewise concordance rather than per-cell significance. What stands is
the direction of the whole pattern; we offer no formal test of it.

\subsection{Transmission in time}\label{sec:transmission}

\begin{figure}[htbp]
\centering
\includegraphics[width=0.62\linewidth]{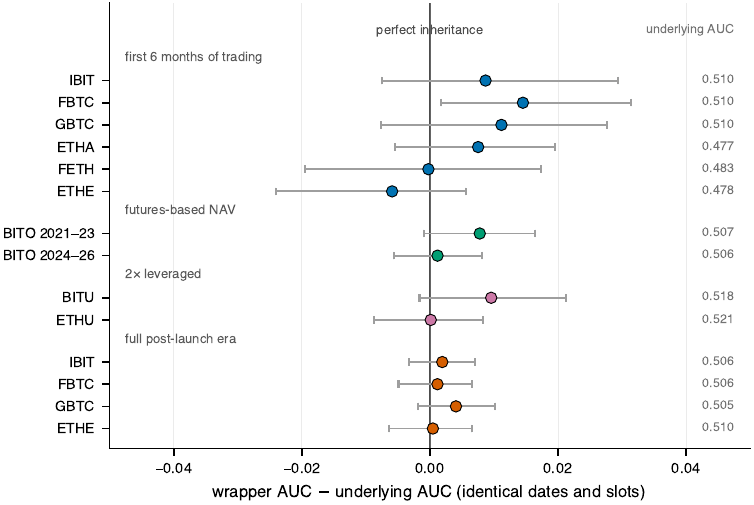}
\caption{Wrapper AUC minus its underlying's on exactly shared bars, with
95\% paired moving-block bootstrap intervals (one block draw applied to
both legs). Zero is perfect inheritance; the right-hand column gives the
underlying's AUC on the shared bars. Per-leg readings:
Table~\ref{tab:events}.}
\label{fig:transmission}
\end{figure}

A cross-sectional concordance still admits a slow story in which listed
venues gradually learn the underlying's dynamics. Three event designs
say otherwise (Figure~\ref{fig:transmission}, Table~\ref{tab:events}).

Bitcoin's three 2024 funds carried the coupling in their first six
months of exchange trading, at fitted values matching their underlying's
($A_w$ between $-0.07$ and $-0.12$ against $-0.08$); the cells are
individually insignificant at ${\sim}1{,}700$ bars, so they establish
concordance, not detection. BITO, priced off CME futures and predating
every spot ETF, matches its underlying over 12{,}813 bars and is the one
era cell that is itself two-model significant at raw 5\% (neither this
era cell nor BITO's panel-span leg survives an FDR correction). The
$2\times$ leveraged BITU and
ETHU reproduce their families' readings through swaps and cash-settled
futures, and the three Ether funds, launching into a window where Ether
itself read below 0.5, reproduce approximately that rather than a
positive signal no process could have handed them.

Formally: thirteen of the fourteen gaps are indistinguishable from zero,
a non-rejection that short launch windows make undemanding; the stronger
statement is that 13 of 14 intervals \emph{exclude} complete
non-inheritance (wrapper AUC $=0.5$), FETH's launch semester being the
exception. Two residuals cut the other way. Twelve of fourteen gaps are
positive (mean $+0.004$; two-sided sign-test $p=0.013$): wrapper tapes
read slightly above the process they wrap, an offset we cannot attribute and the panel cannot exclude; venue-added
structure and wrapper-side tick bounce would both produce it. And FBTC's launch
interval excludes zero ($+0.015$, CI $[+0.002,+0.031]$), about the one
exceedance in fourteen a 5\% interval produces, in the offset's
direction.

\subsection{One-minute timing}\label{sec:leadlag}

Over six months of aligned RTH minutes ($n=49{,}170$), IBIT and Binance
BTC move together at $r=0.982$, while the largest of the twenty
displaced correlations at $\pm1$ to $\pm10$ minutes is $0.013$, about one five-thousandth of the contemporaneous covariation
in $R^2$. There
is no exploitable spot$\to$wrapper lead at this resolution, as expected
if arbitrage enforces the NAV link inside the sampling minute. The
displaced terms are not noise (they alternate in sign and track
Bitcoin's own one-minute autocorrelation), and summed by direction the
wrapper$\to$spot side is the stronger ($+0.015$ against $-0.005$;
difference CI $[-0.028,-0.012]$). We read the asymmetry as IBIT's
transient pricing error mean-reverting within the minute, an
interpretation we have not tested; settling it needs intraday NAV. Its
economic content is bounded either way: the displaced
terms sum to under $0.1\%$ of incremental $R^2$ for the next Bitcoin
minute, far inside any cross-venue round-trip cost. Construction and
diagnostics (alignment placebo, staleness screens, IEX-feed caveats):
Appendix~\ref{app:leadlag}.

The contemporaneity sets the panel's resolution limit. Skipping one bar
between predictors and label severs the contemporaneous echo and leaves
nothing significant anywhere on the session-matched slots (underlying
0.507, largest wrapper 0.497); on the underlying's five-times-longer
24/7 tape the gapped effect survives (BTC 0.514, $p<10^{-3}$), but that
comparison changes the bar set as well as the instrument. A wrapper's
bars carry the underlying's reversal because each bar nearly \emph{is}
the underlying's bar; what survives across the wrapper's own bar
boundary once the link is cut is below this design's resolution, so we
make no claim about a venue margin. Which arbitrage channel binds is
not identified and need not be: all of them transmit the process.

\section{Mechanism evidence: reversal, aggressive order flow, and
capture costs}\label{sec:mech}

Back on the originating tape, three questions remain: what the reversal
travels with, whether the pattern matches compensation for supplying
immediacy, and why nothing has removed it.

\subsection{Reversal follows aggressive flow}\label{sec:generator}

Binance klines record each
bar's taker-buy volume, signing the aggressor side of every trade
without order-book data: the bar's taker imbalance is
$i_t=2V^{\text{taker-buy}}_t/V_t-1\in[-1,1]$. Call a bar
\emph{flow-driven} when its price move agrees in sign with its imbalance
(aggressors pushed the price the way it went; ${\sim}70\%$ of bars) and
\emph{flow-opposed} otherwise. A liquidity-provision account
\citep{grossmanmiller1988,nagel2012}, in which prices concede to absorb
incoming flow and then recover as inventories are worked off, predicts
reversal concentrated after flow-driven moves and increasing in the
flow: the classic volume- and imbalance-conditioned reversal of
equities \citep{campbell1993,chordia2002}, priced in the crypto
cross-section at daily horizons \citep{bianchi2022,farag2025}, here
measured on a tape that signs the aggressor side of every trade.

\begin{figure}[htbp]
\centering
\includegraphics[width=0.84\linewidth]{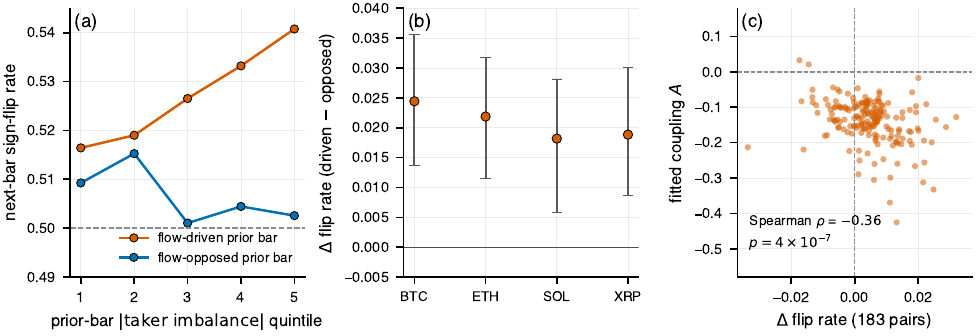}
\caption{Reversal concentrates after flow-driven moves and grows with
the flow. (a)~Pooled next-bar sign-flip rate by quintile of the prior
bar's $|$taker imbalance$|$ (four focal coins, 15\,m, ${\sim}39$k bars
each), split by whether the prior move agreed with its flow: a
dose--response after flow-driven bars, a coin-flip after flow-opposed
ones. (b)~The flow-driven minus flow-opposed flip-rate difference per
coin with moving-block bootstrap CIs (pooled: $+0.021$, CI
$[0.015,0.026]$). (c)~The conditioning scales to the cross-section:
across the 183 pairs, those whose flow-driven moves revert harder carry
systematically stronger fitted reversal couplings (three extreme-$|A|$
pegged/microcap pairs outside the axis range).}
\label{fig:flow}
\end{figure}

Both predictions hold on all four focal coins (Figure~\ref{fig:flow}ab),
and the walk-forward model agrees: its out-of-sample AUC is 0.540 after
flow-driven bars versus 0.519 after flow-opposed ones, so the taker tape
tells you when the coupling will work. The association also holds
across the universe (Figure~\ref{fig:flow}c): the flip-rate difference
retains an independent loading on out-of-sample AUC when volume and
trade size enter jointly, while a retail proxy (median dollar trade
size) carries nothing (Appendix~\ref{app:channels}); what lines up
with the coupling is how hard flow-pushed moves revert, not how small
the customers are.

Flow conditions the reversal but does not replace it
(Table~\ref{tab:subsume}): adding twelve lags of imbalance to the sign
path adds nothing out of sample, while the sign path beats flow alone.
At bar granularity the imbalance is informationally redundant with the
price path it helped create: overreaction plus correction, not a
signal that merely proxies inventory pressure. This is the transitory
price-pressure pattern that \citet{hendershott2014} estimate
structurally in equities, read here at bar level.

Order-book states separate what bar-level flow cannot: whether the
compensation is for absorbing the flow or for rebuilding the depth it
destroyed. Binance publishes ${\sim}30$-second snapshots of cumulative
bid and ask notional within 1--5\% of mid for the USDT-M perpetuals,
where the reversal replicates in full (Appendix~\ref{app:channels}); a
bar's move-side depth consumption is
$\log(D_{\text{open}}/D_{\text{close}})$ of the 1\%-band notional on
the side price moved into. Next-bar reversal is
indistinguishable after depth-consuming and depth-replenished moves:
the pooled flip-rate difference is $-0.002$ (block-bootstrap CI
$[-0.007,+0.004]$), and the contrast stays null at the 2\% band,
residualized on time of day, on open-to-minimum sweep depth, and
inside size-matched $|r_t|$ strata (Appendix~\ref{app:channels}).
Consumption is itself nearly orthogonal to move size and to signed
flow ($|\rho|\le0.09$): within fifteen minutes the book has re-formed,
whatever the move did to it. What conditions the reversal is the
aggressive flow, not any hole it leaves in the book: compensation for
absorbing flow, not for rebuilding depth.

Three
limits bound this reading. The book snapshots are ${\sim}30$-second
states at percentage bands of a moving mid, so consumption finer than
that cadence, and at the touch rather than the band, is unobserved; the
``flow-driven'' label is partly constructed from the contemporaneous
return's own sign, so the split is a conditioning, not a treatment; and
the time-reversal control of Appendix~\ref{app:negcontrols} shows most
of the detectable sign structure is time-symmetric, so the directional
overreaction-and-correction story rests on the flow conditioning here,
not on the sign path itself. Conditioning on the perpetual basis,
funding rate, funding clock, volatility and session is flat in every
case (Appendix~\ref{app:channels}).

One limit is deeper than the rest and bounds the whole section:
ordinary aggressor flow chooses to trade, so whatever makes it push
price may also predict where price goes, and no split of chosen flow
can separate compensated liquidity provision from an informational
account. Separating them needs flow whose initiation carries no
trader's information choice, the intraday analogue of the fire-sale
identification of \citet{coval2007}, and Section~\ref{sec:discussion}
states the corresponding test. Everything in this section is therefore
a conditioning consistent with the provision reading, not a design that
selects it.

\begin{table}[htbp]
\centering
\footnotesize
\begin{tabular}{@{}lcc@{}}
\toprule
Walk-forward feature set & OOS AUC & paired $\Delta$AUC vs.\ signs (CI) \\
\midrule
12 sign lags & 0.529 & --- \\
12 taker-imbalance lags & 0.521 & $-0.008$ $[-0.012,-0.005]$ \\
signs $+$ imbalance & 0.529 & $\phantom{-}0.000$ $[-0.002,+0.002]$ \\
\bottomrule
\end{tabular}
\caption{The subsumption test, pooled across the four focal coins:
bar-level flow neither adds to nor replaces the sign path. Coefficient
detail: Appendix~\ref{app:channels}.}
\label{tab:subsume}
\end{table}

\subsection{The edge is priced inside benchmark transaction
costs}\label{sec:friction}

\begin{figure}[htbp]
\centering
\includegraphics[width=0.73\linewidth]{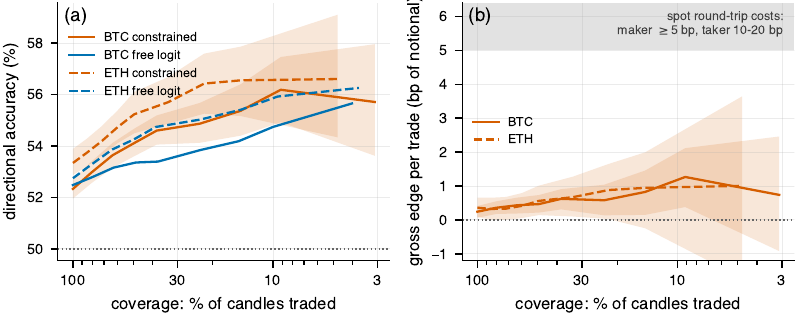}
\caption{The cost of capture (BTC/ETH, 15\,m). (a)~Directional accuracy
as the confidence threshold restricts coverage. (b)~The same thresholds
in return units: the gross edge per trade stays below the shaded spot
round-trip cost band throughout, peaking near 1.3\,bp against a 5\,bp
cheapest band; the 10--20\,bp taker band lies above the axis range.
Bands: 95\% bootstrap CIs.}
\label{fig:cost}
\end{figure}

Why, then, has a measurable signal survived? This subsection prices
the statistical edge against what capturing it would cost
(Figure~\ref{fig:cost}). A deployed rule acts
only when model confidence clears a threshold $|p_t-\tfrac12|\ge\tau$,
with $\tau$ fixed on an ex-ante grid, so nothing is selected on the
out-of-sample series. Raising the threshold does exactly what a real
signal should do to directional accuracy: it rises monotonically as
coverage falls over the range where the trade count supports an
estimate, clearing 56\% on the most confident bars, and rises
more steeply for the constrained model than for the free logit, so
model confidence ranks accuracy rather than merely scaling
it.

The same thresholds in return units tell the economically decisive
half of the story: the gross edge per trade never leaves the low single
digits of a basis point. At its most selective it is about a quarter of
the cheapest realistic spot round-trip band and a tenth of the taker
band, and unthresholded a twentieth of the former. Thresholding
cannot rescue it, because trading fewer, better bars shrinks the
opportunity count faster than it grows the edge. This is a population
statement, not a focal-coin one: not one of the 183 crypto pairs clears
even the 5\,bp maker band at any threshold, the median pair earns
0.46\,bp per trade at $\tau=0.02$ where it trades at all against zero
for the median stock, and moving to 5-minute bars makes matters worse
rather than better (0.15\,bp), because higher frequency shrinks each
opportunity faster than it shrinks costs. (Accuracy
comparisons use the focal coins only, since flat bars inflate accuracy
on thin instruments.) Explicit fees, not the quoted spread, are
what bind here: top-of-book on the focal Binance pairs is of order a
tick, well inside the edge, so a maker who never crosses still pays the
fee schedule (Appendix~\ref{app:data}).

The precise claim is that the edge is \textbf{not exploitable under
benchmark spot-cost assumptions}: a public-schedule taker or maker
paying explicit spot costs cannot capture it anywhere in the
cross-section. That is the statement the limits-to-arbitrage reading
\citep{grossman1980,shleifer1997} needs: a compensation for supplying
immediacy, competed down to roughly the cost of competing for it, is
exactly the residual a costly liquidity-supply equilibrium leaves
standing. It is
not a statement that no trading technology could reach the edge:
sophisticated participants face a cost object we cannot price, and to
the extent they capture part of the compensation they \emph{are} the
liquidity suppliers of Section~\ref{sec:generator}, which reinforces
the account. Which cost component binds this evidence cannot rank; the
gross figures also ignore depth, latency, and adverse selection, all
moving against a would-be taker.

\section{Discussion}\label{sec:discussion}

\paragraph{The account}
At the evidential weights of Section~\ref{sec:intro}: cryptocurrency price processes carry a small, persistent 15-minute mean
reversion, while US equities retain only a residual too weak to rank
direction out of sample; claims listed on those processes carry it too; and
it persists at a size not exploitable under benchmark spot-cost
assumptions \citep{grossman1980,shleifer1997,nagel2012}. The apparent
US-listed ``exceptions'' of Section~\ref{sec:fact}, the metal ETFs, are
then not exceptions but wrappers on crypto-like round-the-clock OTC
processes; the miners, which clear the 24-hour test with momentum-signed
couplings and read null on session-matched slots, are a different
phenomenon and serve as controls without contradiction. The broader
reading, that what orders the classes is intermediation of the primary
market (Table~\ref{tab:main}), is an interpretation the data are
consistent with, not a tested claim: volatility, tick size, and retail
participation order the same classes the same way.

\paragraph{Implications for measuring efficiency}
Two practical corollaries follow. First, the listing venue is the
wrong unit for efficiency measurement whenever the instrument is
priced off an external process: comparisons that pool wrappers with
native-discovery instruments, as most listed universes do,
misattribute the wrappers' readings to the venue unless the NAV
linkage is modeled. Second, measured ``anomalies'' in listed markets
should track what gets wrapped: as ETFs extend into round-the-clock or
lightly intermediated underlyings, their tapes import those processes'
signatures at full size, with no change in the venue's own efficiency.

\paragraph{The equity mirror}
The account also rationalizes the stock side of
Figure~\ref{fig:widedist}: short-horizon reversal was once large in US
equities and shrank as intermediation industrialized
\citep{lehmann1990,jegadeesh1990,chordia2005,chordia2011,nagel2012}. On
our 15-minute bars that history survives as a small negative coupling
in single stocks (Appendix~\ref{app:focal}) that no longer ranks
direction out of sample. Crypto's fragmented, retail-heavy tape
\citep{makarov2020,kogan2024} sits where equities sat before liquidity
provision scaled, and the wrapper evidence shows that listing a claim
on that process inside modern equity structure imports the signature
rather than curing it.

\paragraph{Limitations and falsification}
The wide crypto cross-section is single-exchange and 14 months long,
and even the ex-ante specification conditions on pairs surviving to the
download date (\S\ref{sec:fact}); the focal coins replicate across
venues and years. The six-month holdout is one window: it confirms
survival, not stationarity, and extending it as data accrue costs
nothing but time. The wrapper panel rests on eight underlying families
in one 14-month window, only two with strongly live underlyings on
regular-hours slots; its per-cell, permutation, and resolution limits
are stated where they arise (\S\ref{sec:panelresults},
\S\ref{sec:leadlag}). The mechanism and cost evidence are bounded
as stated in Section~\ref{sec:mech}: bar-level aggressor volume and
${\sim}30$-second band-level book states rather than tick-level order
flow; conditioning rather than identification, so compensated
provision is the account most consistent with the evidence, not one
selected by design; and
gross benchmark costs that do not rank which cost component binds.

Bounded is not unfalsifiable, and each leg is exposed. A NAV-linked
wrapper on a live underlying that fails to inherit, or a correlated
control that robustly does; a wrapper leading its underlying at one
minute by an order of magnitude more than the ${<}0.1\%$ of
incremental $R^2$ we find; reversal
concentrating after flow-\emph{opposed} moves in another market;
reversal after moves pushed by forced liquidation orders (flow that
demands immediacy with no information choice behind its initiation,
observable in public per-order archives) weaker than after matched
ordinary flow-driven moves, which is the informational prediction and
the sharpest discriminating test this design leaves open; a venue
or period in which the gross edge exceeds its own round-trip cost band;
walk-forward AUC on future data converging to 0.5 (the adaptive-markets
reading \citep{lim2011,kristoufek2018}; the first six months of
forward data reject convergence but not erosion); reversal vanishing
on off-grid bar offsets (it does not, Appendix~\ref{app:robust}); or
reversal absent after depth-consuming moves (run: it neither vanishes
nor concentrates, \S\ref{sec:generator}).

\section{Conclusion}\label{sec:conclusion}

Short-horizon directional mean reversion is concentrated in
cryptocurrency markets relative to US equities: a population fact,
measured under one matched out-of-sample protocol, that needs no fitted
model to see and survives every artifact channel we can test, an exact
permutation null, the removal of the common crypto factor, and a frozen
six-month holdout. Its size is a range, not a point: three AUC points as designed, one with
both artifact adjustments applied at once, and every end of the range
clear of zero. US-listed funds priced off
crypto and metal processes carry their underlying's reading, nulls
included, which we read descriptively: a wrapper's tape describes the
process it wraps, not the venue it prints on. What sustains the
reversal, the flow evidence constrains: it concentrates after
aggressive-flow-driven moves, scales with flow intensity, and is
indifferent to the net depth those moves consume, the pattern of
compensated liquidity provision in a book that re-forms within the
bar. The conditioning supports that account but, because chosen flow
may always know something, cannot select it over informational
alternatives, and the edge it leaves is not exploitable under
benchmark spot costs.
Two open tests are sharper than anything run here: the forced-flow
contrast of Section~\ref{sec:discussion}, which would separate
provision from information on flow that cannot choose, and intraday
net asset value for the thin wrappers.

\section*{Funding}
This research received no funding from any agency in the public,
commercial, or not-for-profit sectors.

\section*{Declaration of competing interest}
The authors operate an automated cryptocurrency trading system that
uses a variant of the model studied here; the paper's 15-minute
horizon and up/down label predate this study and mirror that system's
application. No data from the system are used
in this paper, and no design choice reported here was selected on its
results. No other competing interests.

\section*{Data availability}
All market data are obtainable from public sources (Binance, Coinbase,
OKX, Bybit and Dukascopy public APIs; Alpaca market-data API); a
complete replication package is at
\url{https://github.com/nadav2/short-horizon-reversion} (Appendix~\ref{app:repro}).

\section*{Author contributions}
Roles follow the CRediT taxonomy. N.A.K.: Conceptualization,
Methodology, Software, Formal analysis, Investigation, Data curation,
Visualization, Writing -- original draft, Writing -- review \& editing.
J.M.W.: Methodology, Writing -- review \& editing.

\bibliographystyle{abbrvnat}
{\footnotesize
\setlength{\bibsep}{0.1pt plus 0.1ex}
\bibliography{refs}}

\clearpage
\appendix

\section{The measurement device}\label{app:model}

\subsection{Spin transform and model}
Asset returns are heavy-tailed \citep{cont2001,drozdz2018}, so a single
outlier bar can dominate a linear field. Each return is mapped to a
\emph{soft spin} $s_t=\tanh(\lambda r_t)\in(-1,+1)$ with $\lambda=150$
fixed (not fitted); the decision boundary is insensitive to $\lambda$
over $[50,1000]$ (Appendix~\ref{app:robust}). Because $\lambda$ is fixed
while volatilities differ across assets, cross-class comparisons of
coupling \emph{magnitudes} use a volatility-standardized variant
$s_t=\tanh(c\,r_t/\hat\sigma)$, $c=0.45$, with $\hat\sigma$ estimated
causally on the training window; every cross-market conclusion is
unchanged under it. The working model is the kernel-tied logit of
Eq.~\eqref{eq:model}: an $\AR(N)$ logistic regression whose lag weights
are constrained to $w_k=A\,k^{-\alpha}$, leaving three free parameters
$(C,A,\alpha)$ however many lags enter.

\paragraph{Kinetic-Ising correspondence}
Equation~\eqref{eq:model} has the form of the one-step Glauber
transition law \citep{glauber1963} of a kinetic Ising chain with
$J_k=J_0k^{-\alpha}$ couplings \citep{dyson1969}, under $C\equiv2\beta
h_0$, $A\equiv2\beta J_0$ ($\beta$ not separately identifiable). The
correspondence is exact only as $\lambda\to\infty$, where
$s_t\to\sgn(r_t)$; at finite $\lambda$ the model conditions on
$s_t\in(-1,1)$, i.e.\ the naive mean-field closure with $s_t$ a local
magnetization. At $\lambda=150$ the transform is near-linear over the
bulk of 15-minute returns and saturates only above ${\sim}1\%$, so $s_t$
is in practice a soft-winsorized return, only the product $A\lambda$ is
identified in the bulk, and the $\lambda\in[50,1000]$ sweep of
Appendix~\ref{app:robust} is a tail-clipping check, not evidence about
$\lambda$. Unlike Dyson's equilibrium chain the couplings are one-sided in time and antiferromagnetic, with no
detailed balance and no Boltzmann equilibrium; we borrow the coupling form and nothing else. The
multivariate mapping of \citet{campajola2021} requires non-negative
couplings and excludes our regime, but the single-chain correspondence
holds for either sign of $A$. The magnetic vocabulary is retained purely
as labels, and no empirical claim depends on it.

\subsection{Kernel-shape identifiability}\label{app:kernel}
Is the power-law form a finding or a choice? Refitting with three tied
families under the identical protocol (power-law $A\,k^{-\alpha}$,
exponential $A\,e^{-(k-1)/\tau}$, both three parameters, and flat $A$
with two), the \emph{decay} is identified: the flat kernel is worse in
13/15 within-crypto cells, significantly so in 9 (paired block
bootstrap), with large 15\,m deficits (ETH 0.527 vs.\ 0.544). The
functional form is not: power-law and exponential differ by less than
$|0.003|$ AUC in 14/15 cells with sign flips across cells. We therefore
interpret only the existence, sign, and rough decay of the memory
kernel; the fitted $\alpha\approx1$--$1.5$ means ``slow decay over
3--6 bars,'' not scale-free memory.

\subsection{The one-lag baseline}\label{app:nbaseline}
Two short-memory rules are the natural sceptical benchmarks for a
twelve-lag kernel, and we report both rather than leave them implicit:
$N=1$, betting against the previous candle, and $N=3$, the shortest
window at which the ranking gain is already complete. Rerunning the primary
walk-forward with the constrained logit truncated to
$N\in\{1,2,3,6,12\}$, on an identical universe, geometry, spin
transform and selection protocol with only the kernel length changing,
gives Table~\ref{tab:nbaseline}. At $N=1$ the model is
$\sigmoid(C+A s_{t-1})$, the probabilistic form of betting against the
previous candle's sign (the fitted $A$ is negative in 99\% of pairs),
and it already delivers a crypto class mean of 0.5267 (above 0.5 in
96\% of pairs) against the
twelve-lag 0.5305. Because both lag orders are scored on the same
candles the increment is a paired quantity: bootstrapped jointly on the
shared time grid, the crypto class-mean gain from $N=1$ to $N=12$ is
$+0.0039$ (95\% CI $[+0.0026,+0.0052]$, $p<10^{-3}$, positive in 92\% of
pairs): systematic but small, and nearly complete by $N=3$
($+0.0039$ already), consistent with the ${\sim}3$-bar sign
autocorrelation of Figure~\ref{fig:signlag}a. Log-loss and Brier move in
the same direction and are likewise saturated by $N=3$.

Nothing in Sections~\ref{sec:fact}--\ref{sec:mech} turns on the kernel
length: the class contrast is reproduced by the free logit, by $R$, by
DFA, and by the $N=1$ special case alike, with stock class means inside
$[0.4978,0.4993]$ at every $N$. The gap is a crypto-side phenomenon at
every kernel length.

Why twelve, when three ranks as well? $N$ is a measurement window, not a
capacity parameter, and was fixed ex ante rather than read off this
table; the kernel tie makes it nearly free (three parameters at every
$N$), and sweeping $N\in\{6,\dots,48\}$ moves crypto-15\,m AUC by less
than 0.002 while the free $\AR(N)$ logit degrades monotonically. What
the longer window carries is the decay estimate: the share of folds
selecting $\alpha$ at an edge of its search grid falls from 64\% at
$N=2$ to 30\% at $N=12$ (79--56\% on the stock side, where there is no
kernel to find). Ranking saturates by three lags because it ignores
small tail weights; the tail is where the memory is measured.

\begin{table}[htbp]
\centering
\footnotesize
\setlength{\tabcolsep}{6pt}
\begin{tabular}{@{}lcccccccrcc@{}}
\toprule
& \multicolumn{3}{c}{crypto (183)} & \multicolumn{3}{c}{stocks (187)} &
\multicolumn{2}{c}{paired gain (crypto)} & \multicolumn{2}{c}{$\alpha$ (crypto)} \\
\cmidrule(lr){2-4}\cmidrule(lr){5-7}\cmidrule(lr){8-9}\cmidrule(lr){10-11}
$N$ & AUC & log-loss & Brier & AUC & log-loss & Brier &
$\Delta$AUC & \% pairs $>0$ & median & \% edge \\
\midrule
1 & 0.5267 & 0.68238 & 0.24475 & 0.4978 & 0.68106 & 0.24399 & --- & --- & --- & --- \\
2 & 0.5291 & 0.68220 & 0.24466 & 0.4983 & 0.68106 & 0.24399 & $+0.0025$ & 85 & 1.24 & 64 \\
3 & 0.5305 & 0.68209 & 0.24461 & 0.4982 & 0.68107 & 0.24399 & $+0.0039$ & 95 & 1.16 & 51 \\
6 & 0.5303 & 0.68212 & 0.24462 & 0.4987 & 0.68109 & 0.24400 & $+0.0037$ & 91 & 1.23 & 37 \\
12 & \textbf{0.5305} & \textbf{0.68211} & \textbf{0.24462} & 0.4993 &
0.68108 & 0.24400 & $\mathbf{+0.0039}$ & \textbf{92} & 1.27 & \textbf{30} \\
\bottomrule
\end{tabular}
\caption{The constrained logit truncated to $N$ lags, primary universe
and protocol throughout. Columns 8--9: paired crypto class-mean AUC gain
over $N=1$ (joint-bootstrap CI at $N=12$: $[+0.0026,+0.0052]$) and share
of pairs positive. Last two columns: median fitted $\alpha$ and share of
folds selecting $\alpha$ at an edge of its $[0,3]$ grid ($\alpha$ not
free at $N=1$). Betting against the last candle captures most of the
detection; the kernel adds a small increment saturating by $N\approx3$.}
\label{tab:nbaseline}
\end{table}

\subsection{Estimation and baselines}
All models are fit by maximum likelihood: $\alpha$ profiled over a grid
with $(C,A)$ by unpenalized logistic MLE, selected by log-loss on a
chronological validation tail (last 20\% of each training window), then
refit on the full window. The free, ridge, and lasso $\AR(12)$ logits
and the tuned machine-learning baselines select hyperparameters by the
identical protocol, so competitors differ only in hypothesis space.

\subsection{Why the constraint: variance reduction in a weak-signal
regime}\label{app:why}
On a deeper within-crypto dataset (BTC/ETH/SOL/XRP, 5\,m--4\,h, BTC from
2024-03), the three-parameter model beats the unconstrained $\AR(12)$
logit on out-of-sample log-loss in 15/15 cells, with the largest gains
in data-scarce cells: the free logit's train--test gap widens as the AR
order grows while the constrained model's stays near zero.
$L_1$/$L_2$ shrinkage barely moves the free model: norm
penalties do not help a diffuse, ordered signal, whereas the
\emph{shape} prior does. Across folds the free $\AR(24)$ weight vector flips sign at a
given lag in 30\% of folds; the constrained kernel is sign-stable in
100\%. Tuned gradient-boosted trees and an MLP on the same twelve
lagged returns lose to the three-parameter model on out-of-sample
log-loss in all cells, mirroring the shallow-beats-deep pattern in
low-signal financial prediction \citep{gu2020,krauss2017,welch2008}.
Read these against the base rate, not against each other: an
intercept-only predictor scores $\log 2=0.6931$, so on btc-1h the
constrained model (0.6918) is the only one of the four with positive out-of-sample skill; the free
logit (0.6935), GBM (0.6933) and MLP (0.7005) all score worse than predicting the base rate. This is why the
free logit fails log-loss significance even on crypto-15\,m while the
constrained logit clears it: the two agree on AUC, which makes the
contrast estimator-robust, and disagree on calibration, which is what
the constraint buys.

\begin{figure}[htbp]
\centering
\includegraphics[width=0.47\linewidth]{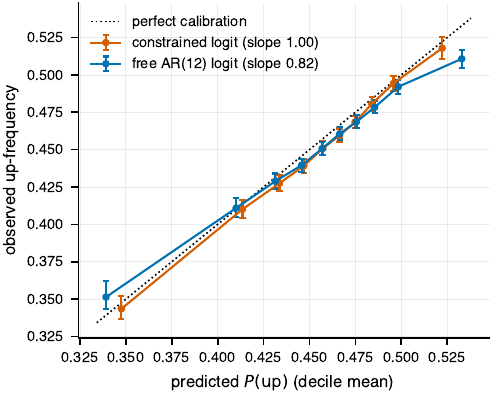}
\caption{Reliability diagram, pooled crypto out-of-sample predictions
(183 pairs): realized up-frequency by decile of predicted
$P(\text{up})$, error bars from the joint block bootstrap. The
constrained logit tracks the diagonal (slope 1.00, CI $[0.95,1.04]$);
the free AR(12) logit over-disperses (slope 0.82, CI $[0.75,0.88]$).}
\label{fig:reliability}
\end{figure}

Figure~\ref{fig:reliability} shows the calibration claim directly
rather than through aggregate score differences. On the pooled crypto
out-of-sample predictions of the primary test (all 183 pairs, 5.1
million bar-level probabilities), the constrained model is calibrated almost perfectly across its own prediction deciles, at a linear
calibration slope of 1.00 (95\% joint block-bootstrap CI $[0.95,1.04]$),
while the free logit over-disperses, with realized frequencies flatter
than predicted at both tails (slope 0.82, CI $[0.75,0.88]$, excluding
1). The pooled log-loss difference, free minus constrained, is $+0.0015$
nats (CI $[+0.0009,+0.0024]$), Brier concordant. Small in absolute
units, but the yardstick is wrong: at AUC 0.531 the binormal
discriminability is $d'=\sqrt2\,\Phi^{-1}(0.531)=0.110$, and the entire
log-loss reduction available to a perfectly calibrated forecaster is
$d'^2/8=0.0015$ nats. The free logit's calibration deficit equals the total extractable
signal: its extra capacity goes into overconfident tails, the failure mode the three-parameter tie removes.

\section{Protocol, inference, and negative controls}\label{app:protocol}

\subsection{Inference details}
The primary test is boxed in Section~\ref{sec:protocol}; it was fixed in
design before the wide universe was run (not a registered pre-analysis
plan). Other horizons reuse the same geometry scaled in bars (1\,h:
2{,}160/360; 4\,h: 720/120). The joint bootstrap that produces the
headline interval uses $B=1{,}000$ resamples (\texttt{dependence});
raising it to $B=5{,}000$ gives $[+0.028,+0.034]$ against the reported
$[+0.027,+0.035]$, and blocks of 192/384/768 fifteen-minute slots at
$B=5{,}000$ give identical conclusions
(\texttt{dependence\_sensitivity}, \texttt{nonflat\_gap}).
The block bootstrap conditions on the fitted models, and two refit
diagnostics quantify what that costs. A per-asset refit-inclusive
bootstrap (50 assets, 25 per class, both models refit inside every
resample, $B=200$) widens intervals by a median factor of 1.12 and
biases the classes in \emph{opposite} directions: crypto $-0.0069$,
stocks $+0.0011$, so the subset's class gap falls from $+0.0278$ to
$+0.0198$ (29\%). Because independent per-asset draws destroy the
cross-sectional dependence, we also ran the joint version
(\texttt{refit\_joint}): one shared block draw per replicate, both
models refit on all 28 assets (chosen at even AUC quantiles per class;
the subset reproduces the universe, gap $+0.0310$ against $+0.0313$).
It gives $+0.0166$ $[+0.0091,+0.0241]$, 54\% retention. The two diagnostics bracket the correction: the fit-conditional headline
is optimistic by a third to a half, putting the refit-corrected gap near
$+0.017$ to $+0.022$. We do not restate the headline on subsets, and
this correction is \emph{not} independent of the artifact adjustments of
Section~\ref{sec:fact}, so the two must not be multiplied. The
refit-inclusive joint SD (0.0038, against the fit-conditional 0.0018)
doubles the interval width, the same message the permutation null below
delivers from another direction.

Block \emph{length} is common to the classes; block \emph{count} is not.
At 384 slots the median crypto pair supplies ${\sim}86$ blocks against
${\sim}16$ for the median stock, and a moving-block bootstrap on 16
blocks has variance-estimator relative error of order
$\sqrt{2/16}\approx35\%$ and under-covering percentile intervals; the per-stock half-widths (0.0185
vs.\ 0.0090 crypto) show it. Rerunning the
per-asset test with block lengths scaled at the standard $n^{1/3}$ rate
(anchored so the median crypto pair keeps 384 slots; the median stock
then resamples ${\sim}28$ blocks of 217 slots) leaves the discovery
counts nearly unchanged: 165 of 183 crypto pairs and the same 5 of
187 stocks after joint BH (\texttt{blocklen\_sensitivity}). The
thin-block problem is therefore one of interval width, not of the
tallies; it still applies untreated to the holdout stock leg
(${\sim}5$ blocks), whose interval should be read accordingly. The
conjunction $p_{\cap}$ is a valid intersection--union $p$-value
\citep{berger1982,bergerhsu1996}, conservative under its null, so the
BH step retains its guarantee.

Our ``$p$'' notation is also doing less than it appears to. The
class-gap $p<10^{-3}$ is the share of observed-centred bootstrap
replicates at or below zero: a percentile-interval statement bounded
below by $1/B$, best read as ``the 99.8\% interval excludes zero'';
the same applies to the DFA gap and the $N{=}1\to N{=}12$ increment.
Per-asset $p$-values use $B=300$ and cannot fall below 0.0033, a floor
that sits near the BH threshold at the first two dozen ranks, so
discovery counts are quantized exactly where marginal discoveries live.

\subsection{An exact permutation null for the class gap}\label{app:perm}
Since neither quantity above is a tail probability under a null, we
construct one, and it is the inference we put first. A single circular
time shift is applied to the label series of every asset simultaneously
on the shared grid, destroying score--label alignment while preserving
within-asset label autocorrelation and the entire cross-sectional
dependence structure. Shifts are whole-day multiples (96 slots): listed
instruments occupy 68 of the 96 slots-of-day against crypto's 96, and a
day-aligned shift maps each asset's coverage onto itself, so no
replicate changes class composition. All 340 admissible shifts are used.

The observed gap lies outside the entire null. Against a null centred at
$-0.0009$ with SD 0.0073 and maximum $+0.0165$, the observed $+0.0313$
sits 4.4 null SDs out, reached by 0 of 340 shifts ($p=0.0029$, the exact
floor); the free logit reads 4.1 SDs, same $p$
(\texttt{permutation\_null}; an unrestricted-shift version agrees to
three decimals). This null assumes no stationarity, symmetry or block
length, and because it preserves the dependence exactly, the population
claim no longer leans on BH-under-dependence. One diagnostic cuts the
other way: the null's dispersion is four times the joint bootstrap's
(0.0073 vs.\ 0.0018). The estimands differ, but the direction matches
the refit widening above, and the reported $[+0.027,+0.035]$ is likely
optimistic in width even though the gap's significance is not in doubt.

\subsection{Power of the equity null}\label{app:power}
The power statement of Section~\ref{sec:fact} is computed entirely from
the frozen per-asset out-of-sample series, with no new modeling. For
each of the 187 stocks the moving-block bootstrap (block 384 slots, as
in the per-asset test) gives the sampling dispersion
$\hat\sigma_i$ of its out-of-sample AUC; under a normal approximation of
the bootstrap law, the power of the one-sided 5\% test against a true
excess $\Delta$ is $\Phi(\Delta/\hat\sigma_i - z_{0.95})$. At
$\Delta=0.031$ (the crypto class mean) the per-stock powers are as
quoted in Section~\ref{sec:fact}, with power $\ge 0.9$ in 78\% of
stocks (the weakest series has $n_{\text{OOS}}\approx3{,}000$); the
per-asset minimum excess detectable with 99\% power is 0.033 at the
median stock, which is why the class-level bound, whose power comes
from pooling, carries the claim. The joint moving-block bootstrap of the
stock class-mean AUC on the shared time grid (as in the primary gap
inference) gives an observed mean of 0.499 with the one-sided 95\% upper
bound of 0.5007 quoted there, excluding any common equity-class effect
larger than ${\approx}0.0007$ of AUC (\texttt{power}).

\subsection{Negative controls}\label{app:negcontrols}
\emph{Shuffled labels} destroy all feature--label dependence: on the
focal panel, mean AUC 0.500 and 0/14 significant; at full scale
(all 370 series through the identical pipeline, FDR step included), mean
AUC 0.499 (crypto) and 0.496 (stocks), class gap $+0.002$ against the
observed $+0.031$, and
0/370 significant after FDR. The machinery cannot manufacture the
result. \emph{Phase-randomized surrogates} (IAAFT)
\citep{theiler1992,schreiber1996} preserve linear autocovariance while
destroying nonlinear dependence. On the focal coins the signal collapses
(BTC $0.533\to0.506$), pointing to a predominantly nonlinear signal; BTC's $\rho_1=-0.003$
supports only 0.501 through the linear channel of
\S\ref{sec:fact} against 0.533 observed. But the control itself is
flawed, in three ways we report rather than smooth over. It is
anti-conservative on session-heteroskedastic series: surrogates clear
the two-model criterion on 7 of 14 instruments (5 of 11 non-crypto
against 3 of 11 real), because IAAFT preserves $\rho_1$ while destroying
the volatility clustering that makes it unrankable (SPX:
$\rho_1=-0.030$, real AUC 0.492, variance-homogenized 0.515); the free
logit reaches a mean AUC of 0.520 on surrogates, bounding how much of
the conjunction criterion's second leg is linear read-out. And at class
level the arithmetic does not reconcile: the crypto class-mean
$\rho_1=-0.046$ implies a linear-channel AUC of 0.521, two thirds of the excess, yet
surrogates retain a fifth. The per-asset accounting
resolves that discrepancy as compositional: across the 183 pairs, each
asset's linear-implied excess $\Phi(\sqrt{4/\pi}\,|\rho_1|)-0.5$ and its
observed excess are nearly uncorrelated ($r=+0.06$; median crypto
$\rho_1$ is $-0.028$ against the mean's $-0.046$), so the class-mean
$\rho_1$ is carried by bounce-prone pairs whose AUC excess is no larger,
and a class-mean linear implication does not describe any particular
pair (\texttt{linear\_channel}). The mirror image holds for stocks:
their linear channel implies $+0.015$ of excess that never ranks
(observed $-0.001$), the heteroskedasticity reading in cross-section.
The IAAFT row bounds the linear channel and certifies nothing.

The null with correct size is the sign-randomization surrogate: hold each series' $|r|$ path fixed, so volatility clustering and the
diurnal profile survive exactly, and randomize only the signs, iid at
the empirical up-rate, destroying precisely the dependence under test.
Run on the fourteen focal instruments under the identical walk-forward and bootstrap, it is clean: \textbf{0 of 14} clear the
two-model criterion (crypto mean AUC 0.496, non-crypto 0.498), against
IAAFT's 7 of 14 (\texttt{signrand}). The machinery finds nothing once
sign dependence is removed, with every magnitude structure intact.

\emph{Time reversal}, a third control, preserves the autocovariance
exactly while destroying time-\emph{asymmetric} structure, and it cuts
against one of our readings. The crypto mean falls only 0.536 to 0.521
with all three coins still two-model significant and couplings still
negative; the non-crypto side reads 0.501, 0 of 11. So the class contrast is robust to reversal, one more estimator-
independent confirmation, but most of the detectable structure is time-
symmetric,
and the directional overreaction-then-correction story of
Section~\ref{sec:mech} therefore rests on the flow conditioning, not on
the sign path.

A generative check: a Glauber chain simulated from the fitted parameters
reproduces the predictive level (AUC 0.535 vs.\ 0.538) but over-produces
linear signatures ($\rho_1$ to $-0.07$ against an observed
${\approx}-0.003$, a ${>}10$-SE rejection at $n\approx33{,}000$;
$\VR(16)\approx0.84$ vs.\ 0.97). It is an adequate predictive reduction and an incomplete generative
model. That is unsurprising: Eq.~\eqref{eq:model} consumes a magnitude
and emits a sign probability,
so any simulation must supply a sign-to-return map the estimator never
specified.

\section{Focal panel and horizon dependence}\label{app:focal}

Under the two-model criterion the focal tally is crypto 3/3, FX 2/3,
US-listed 1/8 (Table~\ref{tab:main}); permuting class labels over
per-instrument AUCs, the crypto-minus-rest gap of $+0.030$ has
$p=0.003$. AAPL and NVDA carry the familiar small negative
bid--ask-bounce coupling of individual equities
\citep{roll1984,lehmann1990,lomackinlay1990} yet barely clear AUC 0.5;
crypto's coupling is two-to-three times larger and strong enough to rank
direction out of sample. Focal mean AUC by horizon
(Figure~\ref{fig:focal}): crypto 0.536/0.521/0.496 at 15\,m/1\,h/4\,h,
FX 0.513/0.507/0.487, US-listed ${\approx}0.50$ throughout. By the
bootstrap, crypto is significant in 4/4 coins at 1\,h under the
constrained logit with the free logit corroborating 3/4 (SOL from the
independent-venue refetch, as at 15\,m); FX drops to
0/3 at 1\,h; US-listed is 0/8 at both 1\,h and 4\,h. At 4\,h
$n_{\text{OOS}}\approx10^3$ and the apparent crypto-reversal/
equity-momentum flip is suggestive only. Crypto coupling stays negative
at 1\,h ($\bar A\approx-0.13$) before fading toward zero at 4\,h;
US-listed drifts from weak reversal to weak momentum.

\section{Robustness battery}\label{app:robust}

\paragraph{Hyperparameters and standardized spins}
Sweeping $\lambda\in\{50,\dots,1000\}$ and $N\in\{6,\dots,48\}$ moves
mean crypto-15\,m AUC by less than 0.002. The volatility-standardized
spin leaves the picture unchanged (crypto 0.535, $\bar A=-0.32$;
non-crypto 0.505, $\bar A=-0.07$) and shrinks the visually inflated
fixed-$\lambda$ FX couplings of Table~\ref{tab:main} to $-0.14$; coupling
magnitudes should be compared on the standardized scale.

\paragraph{Regular trading hours}
Restricting the eight US-listed focal instruments to 09:30--16:00 New
York yields mean AUC 0.500 with 0/8 significant (gold's marginal 24-h
significance disappears); the full 187-stock universe on RTH bars gives
mean 0.498 with 1.1\% significant. Crypto, having no session, is
unchanged: on like-for-like active-session bars the contrast widens
slightly.

\paragraph{Multi-year stability}
Extending the four focal coins to 2021: all 24 coin-years significant
under both models (AUC 0.523--0.550), couplings negative in every cell,
with a mild downward BTC drift consistent with adaptive-markets readings
\citep{kristoufek2018,khuntia2018,lim2011}
(Figure~\ref{fig:stability} in the main text). With nothing fitted, the
sign correlation $R$ per calendar quarter is negative in all 21 crypto
quarters and 83/84 coin-quarters, while the US-listed mean never leaves
$[-0.020,+0.007]$. The equity counterpart (40 instrument-years,
including 2022) yields mean AUC 0.501 with ${\approx}$chance-level
isolated significances. Crypto AUC is also flat across trailing
volatility terciles (0.536/0.539/0.534) and UTC sessions
(0.531--0.542): a standing feature of the tape, not an episodic one.

\paragraph{Detrended fluctuation analysis}
As an estimator outside the conditional-model family, DFA-1
\citep{peng1994,kantelhardt2002} on all 370 return series (twelve
log-spaced scales, 8--512 candles) gives crypto a class-mean scaling
exponent of $H=0.469$ (below $1/2$ in 88.5\% of pairs) against 0.510 for
US-listed instruments (28.3\%), the model-free counterpart of a negative
coupling \citep{hurst1951}; under the joint bootstrap (with $H$
re-estimated on block-capped scales inside each replicate) the class gap
is $-0.049$ $[-0.084,-0.012]$ (\texttt{dfa}).
Three caveats keep this row modest. It is not an independent probe:
negative short-lag autocorrelation mechanically depresses DFA-1 $H$, so
antipersistence and a negative coupling summarize one autocovariance
structure. The estimate is scale-range sensitive (full-sample gap
$-0.041$ against block-capped $-0.049$, a swing comparable to the
interval's distance from zero), and these near-martingale series need
carry no scaling regime at all, so we attach no meaning to the level.
And one day is 96 candles, inside the fitting range: session periodicity
puts a crossover in the stock fluctuation functions that 24/7 crypto
lacks, biasing a single-slope fit asymmetrically, in the direction that
widens the gap. Rerunning the stock leg on session-restricted bars
(09:30--16:00 weekdays, the overnight cycle spliced out) confirms the
bias and sizes it: the stock class mean falls from 0.510 to 0.495, with
the below-$1/2$ share rising from 28\% to 61\% (\texttt{dfa\_rth}). So
roughly a third of the 24-hour class gap was session periodicity, the
like-for-like gap is nearer $-0.026$ than $-0.041$, and stocks are themselves weakly antipersistent intraday, consistent with
their small negative bounce $\rho_1$. The ordering survives (crypto 0.469
remains below the session-restricted stock mean), and the row
corroborates the ordering established by $R$, no more.

\begin{figure}[htbp]
\centering
\includegraphics[width=0.82\linewidth]{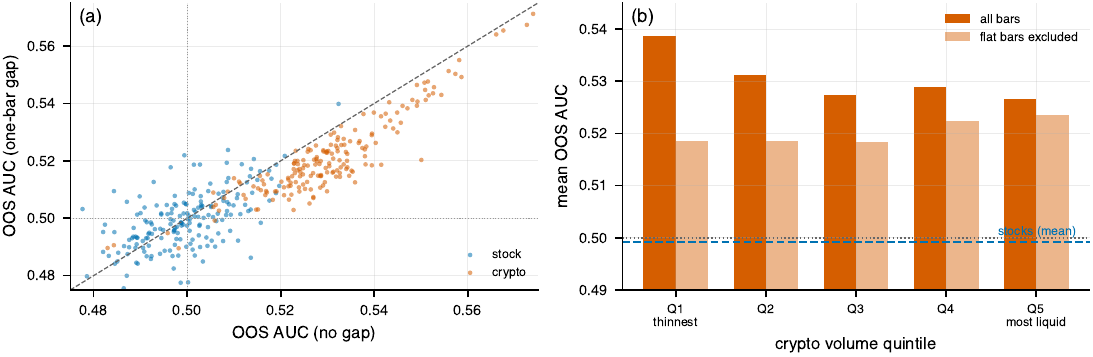}
\caption{Microstructure-artifact tests at population scale.
(a)~Per-asset AUC with vs.\ without a one-bar gap between features and
label. (b)~Crypto AUC by volume quintile, all bars vs.\ non-flat bars:
the thin-pair advantage is a flat-bar artifact; the artifact-free
signal is uniform. Dashed line: stock mean; dotted line: 0.500.}
\label{fig:micro}
\end{figure}

\paragraph{Microstructure artifacts}
The one-bar-gap test (predicting bar $t$ from $t-2,\dots,t-13$) retains
BTC 0.514, ETH 0.519, XRP 0.513 (all $p<10^{-3}$ both models); at
population scale crypto moves 0.531$\to$0.522 with 81\% of pairs still
significant, stocks 0.499 either way (Figure~\ref{fig:micro}a).
Excluding flat bars (median 16\% of bars across wide US-listed
instruments, 6\% wide crypto, $<1\%$ focal coins) moves wide crypto
0.531$\to$0.520 and stocks 0.4993$\to$0.4985, with 97\% of pairs above
0.5. This is the preferred artifact-robust gap of Section~\ref{sec:fact},
whose interval is computed under the identical joint bootstrap at
$B=5{,}000$ ($p<2\times10^{-4}$; \texttt{nonflat\_gap}). Accuracy
metrics are far more sensitive to the flat-bar convention, which is why
raw accuracy is never used as a headline metric. Stratifying by volume, the modest raw AUC--liquidity gradient is
carried entirely by the flat-bar channel (non-flat quintile means
0.518--0.524, flat; Figure~\ref{fig:micro}b); regressing
flat-bar-robust AUC on log volume, flat share, and median $|r_t|$ (HC1),
the intercept is 0.520 and the volume slope slightly positive, the
opposite of a staleness artifact. (The raw ex-ante volume--AUC rank
correlation of $-0.16$ and this conditional slope differ in sign;
both are economically zero, and the ledger's ``volume slope
${\approx}0$'' refers to this conditional specification.)

\paragraph{Bar-grid alignment}
Quarter-hour clock effects, meaning order-flow or quoting seasonality
concentrated at :00/\allowbreak:15/\allowbreak:30/\allowbreak:45, could
in principle manufacture
structure specific to the conventional 15-minute grid. Rebuilding bars
from public 1-second klines with openings shifted to $+1$, $+2$, $+5$,
and $+7$ minutes past the quarter-hour, the sign correlation $R$ of
Eq.~\eqref{eq:R} stays strongly negative at every offset on the three
focal coins with 1-second archives (SOL lacks one): BTC $-0.063$
on-grid against $-0.047$ to $-0.061$ off-grid; ETH $-0.094$ against
$-0.058$ to $-0.073$; XRP $-0.075$ against $-0.050$ to $-0.073$; all
fifteen 95\% block-bootstrap intervals exclude zero, and the on-grid row
is rebuilt from that same archive under the same bar-completeness rules
as the off-grid rows (2026-02--2026-05, ${\sim}9{,}400$ bars per offset;
${\sim}4{,}400$ for XRP, whose archive ends 2026-03).
The signal does not depend on the conventional grid; the on-grid
values are the largest in magnitude (most visibly for ETH), consistent
with a small clock-aligned component on top of it. The check covers the
focal coins, the instruments with public 1-second archives.

\paragraph{External validity}
Refetching the four focal coins from Coinbase, OKX, and Bybit and
re-running everything: all 12 venue--coin cells significant under both
models with per-coin AUC within $\pm0.002$ of Binance. Dukascopy
bid/ask-averaged quote-midprice bars, on which bid--ask bounce cannot
exist, reproduce the effect (BTC 0.529, ETH 0.537, couplings
${\approx}-0.30$). Those bars carry a median quoted spread of 5.4\,bp of
mid for BTC and 17\,bp for ETH against a thresholded gross edge of
${\sim}1$\,bp, but Dukascopy quotes a retail contract-for-difference,
not the Binance book, where top-of-book on these pairs is of order a
tick. We therefore use them for the bounce control rather than
as an independent price of the friction, which is set by fees
(Appendix~\ref{app:data}). Close-to-close labels coincide with open-to-close
to the third decimal; a VWAP-to-VWAP variant scores higher only through
the mechanical smoothing dependence of time-averaged prices
\citep{working1960} and is not tradable structure.

\paragraph{Universe selection}
The 183-pair universe was frozen on a post-sample volume ranking
(survivor set; Appendix~\ref{app:data}). The ex-ante specification of
Section~\ref{sec:fact}, whose headline statistics are reported there,
is recomputed by aggregation over the frozen per-asset outputs with no
model refits (\texttt{exante\_primary}); the counts behind the
percentages are 129/133 vs.\ 5/187, the within-class mean pairwise
correlation is 0.54 ($N_{\text{eff}}\approx1.9$), and the ex-ante
top-100 subset gives a gap of $+0.033$ $[+0.029,+0.037]$. Across
ex-ante volume quintiles crypto mean AUC is flat (0.530--0.538).

\section{Wrapper panel: instrument-level detail}\label{app:panel}

\begin{table}[htbp]
\centering
\footnotesize
\setlength{\tabcolsep}{4pt}
\begin{tabular}{@{}lllccccr@{}}
\toprule
Underlying & Instrument & Role & AUC & non-flat & 95\% CI & conj.\ $p$ & $A$ \\
\midrule
Gold & XAUUSD & underlying & 0.503 & 0.503 & [0.489, 0.518] & 0.282 & $-0.28$ \\
 & GLD & wrapper & 0.507 & 0.506 & [0.491, 0.520] & 0.233 & $-0.20$ \\
 & GDX & correlated & 0.499 & 0.500 & [0.483, 0.512] & 0.605 & $-0.01$ \\
 & NEM & correlated & 0.501 & 0.500 & [0.484, 0.515] & 0.704 & $-0.02$ \\
\addlinespace[2pt]
Silver & XAGUSD & underlying & 0.500 & 0.500 & [0.482, 0.515] & 0.546 & $-0.03$ \\
 & SLV & wrapper & 0.505 & 0.500 & [0.488, 0.521] & 0.276 & $-0.03$ \\
\addlinespace[2pt]
Platinum & XPTUSD & underlying & \textbf{0.532} & 0.531 & [0.517, 0.548] & \textbf{0.001} & $-0.30$ \\
 & PPLT & wrapper (thin) & 0.511 & 0.505 & [0.492, 0.528] & 0.129 & $-0.06$ \\
\addlinespace[2pt]
Palladium & XPDUSD & underlying & \textbf{0.523} & 0.521 & [0.507, 0.536] & \textbf{0.006} & $-0.21$ \\
 & PALL & wrapper (thin) & 0.510 & 0.502 & [0.492, 0.526] & 0.157 & $-0.10$ \\
\addlinespace[2pt]
Bitcoin & BTC & underlying & \textbf{0.524} & 0.525 & [0.513, 0.538] & \textbf{0.025} & $-0.19$ \\
 & IBIT & wrapper & \textbf{0.525} & 0.524 & [0.511, 0.540] & \textbf{0.004} & $-0.20$ \\
 & FBTC & wrapper & \textbf{0.525} & 0.525 & [0.513, 0.541] & \textbf{0.008} & $-0.24$ \\
 & BITB & wrapper & \textbf{0.525} & 0.526 & [0.512, 0.540] & \textbf{0.002} & $-0.24$ \\
 & ARKB & wrapper & \textbf{0.523} & 0.523 & [0.509, 0.537] & \textbf{0.009} & $-0.22$ \\
 & HODL & wrapper & \textbf{0.533} & 0.532 & [0.519, 0.546] & \textbf{0.000} & $-0.25$ \\
 & BTCO & wrapper & \textbf{0.516} & 0.507 & [0.502, 0.532] & \textbf{0.017} & $-0.17$ \\
 & EZBC & wrapper & \textbf{0.517} & 0.514 & [0.501, 0.532] & \textbf{0.026} & $-0.22$ \\
 & GBTC & wrapper (conv.\ trust) & \textbf{0.521} & 0.520 & [0.507, 0.536] & \textbf{0.005} & $-0.19$ \\
 & BITO & futures NAV & \textbf{0.524} & 0.523 & [0.512, 0.537] & \textbf{0.030} & $-0.23$ \\
 & BITU & $2\times$ lev.\ (deriv.) & \textbf{0.527} & 0.527 & [0.514, 0.542] & \textbf{0.004} & $-0.15$ \\
 & COIN & correlated & 0.520 & 0.520 & [0.506, 0.535] & 0.364 & $-0.11$ \\
 & MSTR & correlated & 0.506 & 0.505 & [0.490, 0.521] & 0.248 & $-0.05$ \\
\addlinespace[2pt]
Ether & ETH & underlying & 0.524 & 0.523 & [0.507, 0.538] & 0.150 & $-0.17$ \\
 & ETHA & wrapper & 0.522 & 0.522 & [0.508, 0.539] & 0.407 & $-0.16$ \\
 & FETH & wrapper & 0.522 & 0.522 & [0.506, 0.536] & 0.398 & $-0.17$ \\
 & ETHE & wrapper (conv.\ trust) & 0.525 & 0.524 & [0.510, 0.539] & 0.345 & $-0.16$ \\
 & ETHU & $2\times$ lev.\ (deriv.) & 0.522 & 0.522 & [0.508, 0.537] & 0.378 & $-0.12$ \\
\addlinespace[2pt]
Euro & EURUSD & underlying & 0.504 & 0.504 & [0.489, 0.518] & 0.263 & $-0.57$ \\
 & FXE & wrapper (thin) & 0.484 & 0.493 & [0.467, 0.501] & 0.967 & $-0.25$ \\
 & UUP & correlated & 0.503 & 0.502 & [0.489, 0.516] & 0.472 & $-0.04$ \\
\addlinespace[2pt]
Yen & USDJPY & underlying & 0.508 & 0.507 & [0.494, 0.522] & 0.148 & $+0.05$ \\
 & FXY & wrapper (thin, inv.) & 0.499 & 0.495 & [0.486, 0.511] & 0.638 & $+0.14$ \\
\bottomrule
\end{tabular}
\caption{Every panel leg on session-matched RTH bars (identical NY
09:30--16:00 slots, 2025-01-01 to 2026-02-11; identical walk-forward,
models, bootstrap). ``Non-flat'' rescores with flat bars dropped; CI,
$p$, $A$ refer to the all-bars fit. Bold: two-model significant at raw
5\% (within-panel FDR: \S\ref{sec:panelresults}). ``Thin'': $>15\%$ flat
bars on the 24-hour tape (2.7--10.7\% on these RTH slots). ``Inv.'':
FXY moves inversely to the USDJPY quote; AUC and the sign of $A$ are
invariant under $r\to-r$.}
\label{tab:panel}
\end{table}

Rescoring on non-flat bars moves ten of the seventeen wrappers by less
than $0.001$, and the exceptions are informative: the two thin metal
wrappers fall further (PPLT $0.511\to0.505$, PALL $0.510\to0.502$), as
do two of the eight Bitcoin funds (BTCO $0.516\to0.507$, EZBC
$0.517\to0.514$), while FXE rises towards the band
($0.484\to0.493$). The family readings of Table~\ref{tab:family}
survive it: the Bitcoin wrapper mean moves to 0.522 against an
underlying of 0.525 (family gap $-0.003$ against $-0.001$) and the
Ether mean to 0.523 against 0.523 ($-0.001$ against $-0.000$). On
24-hour bars, where their full sessions enter, all four spot metals are
themselves two-model significant (AUC 0.513--0.522, $A=-0.18$ to
$-0.24$): the OTC-metals analogue of the crypto finding, sitting
between FX and crypto.

Table~\ref{tab:panel} gives every panel leg; the twelve event-study era
cells are in Table~\ref{tab:events} below.

\begin{table}[htbp]
\centering
\footnotesize
\setlength{\tabcolsep}{5pt}
\begin{tabular}{@{}llllcccc@{}}
\toprule
Era & Span & $n_{\text{OOS}}$ & flat & wrapper AUC & underl.\ AUC & gap & $A_w$ / $A_u$ \\
\midrule
IBIT first 6 mo & 2024-01\,--\,2024-07 & 1{,}690 & 1.1\% & 0.519 & 0.514 & $+0.005$ & $-0.07$ / $-0.08$ \\
FBTC first 6 mo & 2024-01\,--\,2024-07 & 1{,}690 & 1.0\% & 0.525 & 0.514 & $+0.010$ & $-0.12$ / $-0.08$ \\
GBTC first 6 mo & 2024-01\,--\,2024-07 & 1{,}690 & 1.3\% & 0.521 & 0.514 & $+0.007$ & $-0.08$ / $-0.08$ \\
ETHA first 6 mo & 2024-07\,--\,2025-01 & 1{,}742 & 2.3\% & 0.485 & 0.486 & $-0.001$ & $-0.05$ / $-0.01$ \\
FETH first 6 mo & 2024-07\,--\,2025-01 & 1{,}737 & 2.4\% & 0.483 & 0.486 & $-0.003$ & $-0.06$ / $-0.01$ \\
ETHE first 6 mo & 2024-07\,--\,2025-01 & 1{,}740 & 2.4\% & 0.472 & 0.486 & $-0.014$ & $-0.04$ / $-0.01$ \\
\addlinespace[2pt]
BITO 2021--23 & 2021-10\,--\,2023-12 & 12{,}813 & 4.0\% & \textbf{0.515} & 0.514 & $+0.001$ & $-0.15$ / $-0.14$ \\
BITO 2024--26 & 2024-01\,--\,2026-02 & 12{,}038 & 2.3\% & 0.507 & 0.508 & $-0.001$ & $-0.11$ / $-0.11$ \\
IBIT full era & 2024-01\,--\,2026-02 & 12{,}038 & 1.0\% & 0.508 & 0.508 & $+0.000$ & $-0.11$ / $-0.11$ \\
FBTC full era & 2024-01\,--\,2026-02 & 12{,}034 & 0.9\% & 0.507 & 0.508 & $-0.000$ & $-0.12$ / $-0.11$ \\
GBTC full era & 2024-01\,--\,2026-02 & 12{,}024 & 1.1\% & 0.509 & 0.508 & $+0.002$ & $-0.12$ / $-0.11$ \\
ETHE full era & 2024-07\,--\,2026-02 & 8{,}595 & 2.2\% & 0.511 & 0.508 & $+0.002$ & $-0.12$ / $-0.12$ \\
\bottomrule
\end{tabular}
\caption{Wrapper event studies on session-matched RTH bars, each leg
scored on its own bars (unpaired gaps; paired:
Figure~\ref{fig:transmission}). $n_{\text{OOS}}$ and ``flat'' describe
the wrapper leg; $A_w$/$A_u$ are fitted couplings. Bold: two-model
significant at raw 5\% (BITO 2021--23, conj.\ $p=0.042$). Alpaca's feed
begins at GBTC/ETHE's uplisting, so their trust eras are unobservable.}
\label{tab:events}
\end{table}

\subsection{One-minute construction and diagnostics}\label{app:leadlag}

\paragraph{Construction}
Both one-minute series carry bucket-start UTC timestamps (verified
empirically: the bar labelled 09:30 New York has the opening minute's
elevated variance), and bars are paired only on exactly equal epoch
seconds, with no nearest-neighbor joining. Returns are intra-bar
open-to-close, so consecutive bars do not overlap and the inter-bar gap
is discarded; this attenuates rather than inflates displaced
correlations, and it excludes the one place spot mechanically leads the
ETF, the accumulated overnight move at the 09:30 open. Lagged pairs are
formed only within runs of consecutive minutes, so no lag straddles a
session boundary. The session filter excludes weekends, non-trading
days (by intersection), and the post-13:00 minutes of the two NYSE
early-close days in span; an earlier version admitted those 306
post-close minutes, and excluding them \emph{raises} the contemporaneous
correlation slightly. $n=49{,}170$ aligned minutes in 127 daily runs.

\paragraph{Diagnostics}
Three checks on the headline numbers. \emph{Alignment placebo:}
deliberately shifting the IBIT series by $\pm1$ minute before the join
collapses the lag-0 correlation from $0.982$ to ${\approx}0$ and
relocates the peak to exactly $k=\pm1$; the machinery resolves a
one-minute offset, so the lag-0 peak is not an artifact of coarse
joining. \emph{Authenticity:} the two series are not accidental copies
of one another, since the IBIT returns lie on a half-penny price grid whose
implied price tracks IBIT's actual 2025--26 path, and the regression
beta of IBIT on BTC is 1.006. \emph{Staleness:} our archived one-minute
records retain bar open and close only, so zero return is the staleness
proxy; 4.3\% of aligned IBIT minutes have exactly zero return, and
excluding them moves the contemporaneous correlation to $0.983$ and the
largest displaced term to $0.013$, changing nothing. Alpaca's bars also
carry per-bar volume and trade count, which would give a sharper screen;
given how little the zero-return exclusion moves the numbers we did not
rerun on it.

\paragraph{Feed caveats}
The US leg is Alpaca's IEX feed, not consolidated SIP/NBBO data: IBIT's
09:30 bar opens at the first IEX print rather than the official auction
open, the close misses the closing cross (the per-minute correlation
degrades from 0.981 midday to ${\approx}0.95$ in the final minute), and
minutes with no IEX print are absent rather than zero-filled, so the
sample is IEX-active minutes. All of these degrade the measured
contemporaneity, so the reported $r=0.982$ is, if anything, a lower
bound on the consolidated-tape value; a replication on SIP data is the
natural upgrade and requires only a data purchase, not a design change.

\section{Mechanism details: subsumption and the flat
channels}\label{app:channels}

\paragraph{Subsumption}
The pooled feature-set comparison is Table~\ref{tab:subsume} of the
main text. Behind it, the lag-1 sign coefficient attenuates by roughly
a quarter when imbalance enters and remains dominant. In the 183-pair
cross-section,
$\Delta$flip retains an independent loading on out-of-sample AUC
($+0.004$ per SD, $t=3.0$, HC1) when dollar volume and median trade
size enter jointly; trade size itself carries nothing ($t=1.0$).
HC1 treats the 183 pairs as independent draws, which at mean pairwise
return correlation 0.40 they are not, so these $t$-statistics are
optimistic in level; we read the sign and the ordering, not the
significance.

\paragraph{Book states}
Per-coin flip-rate differences after depth-consuming versus
depth-replenished moves: BTC $+0.002$, ETH $-0.009$, SOL $+0.001$, XRP
$-0.001$, every block-bootstrap CI straddling zero; pooled $-0.002$
$[-0.007,+0.004]$, size-matched within $|r_t|$ quintiles $-0.0005$
$[-0.006,+0.005]$, dose-response slope flat. The null repeats at the
2\% band, residualized on the 96 intraday slots, and with
open-to-minimum sweep depth; walk-forward AUC is likewise flat across
prior-bar consumption terciles (0.537, 0.531, 0.538). Two book-state
gradients appear and neither disturbs the account: flip rates are
mildly higher on a deep pre-bar book (time-of-day confounded) and rise
monotonically in $|r_t|$ over pre-bar depth. Consumption correlates
with neither $|r_t|$ nor signed imbalance ($|\rho|\le0.09$): the
fast-replenishment point in correlation form.

\paragraph{Channels that are flat}
\emph{Perp--spot arbitrage:} the reversal replicates fully on the
perpetual-futures tape (AUC 0.531--0.538, $A=-0.21$ to $-0.36$) but is
flat in the lagged $|$basis$|$ (pooled tercile AUCs 0.533, 0.537,
0.536). If basis-closing flows generated it, it should strengthen when
the basis is stretched. \emph{Funding carry:} flat in the lagged funding rate
(0.532, 0.537, 0.537); predictions in the hour containing a funding
settlement are slightly \emph{less} accurate (0.531 vs.\ 0.537).
\emph{Volatility:} flat across trailing-volatility terciles.
\emph{Session:} the edge peaks in the most active hours (10:00--14:00
and 21:00 UTC) and sags through the Asian morning, so the signal lives
on the active tape rather than in empty sessions.

\section{Data construction and universe details}\label{app:data}

\paragraph{Conventions}
All bars are stamped UTC with bucket-start timestamps on a shared grid;
missing buckets are left absent, never filled. Labels are intra-bar
open-to-close, so gaps, splits, and dividends never cross a bar
boundary. Flat bars are retained (flat-bar-robust rescoring throughout;
Appendix~\ref{app:robust}).

\paragraph{Crypto (Binance)}
Spot klines from the public REST API; Binance emits a bucket for every
grid slot, including ones with no trades, so the bar grid is complete
and no bar is imputed. Flat bars are therefore real intervals, but on
a universe of the highest-volume pairs they are almost entirely bars
that traded and whose net move rounded to zero at the tick, not
no-trade intervals, which is why the flat-bar channel is a labelling
artifact rather than a liquidity measure. Taker imbalance uses the kline taker-buy base-volume field;
perpetual legs from the USDT-M futures API; funding history from the
funding-rate endpoint. Order-book depth is the public Binance Vision
\texttt{bookDepth} archive for the USDT-M perpetuals, ${\sim}30$-second
snapshots of cumulative notional within 1--5\% of mid, covering
${\ge}99\%$ of panel-span bars (one missing day, XRP 2026-01-14).
The wide universe was selected on 2026-06-08
(post-sample) as the highest 24-hour quote-volume USDT spot pairs,
excluding pegged-base pairs and leveraged tokens, requiring
$\ge$5{,}000 bars in span: a survivor universe, with the stock list
curated at the same date so the cross-class contrast is symmetric;
Appendix~\ref{app:robust} bounds the selection channels empirically.

\paragraph{Post-sample holdout accounting}
The holdout refetch of Section~\ref{sec:fact} covers 2026-02-12 to
2026-08-08 with relaxed minimum-bar cuts, disclosed in the output file:
4 of the 183 frozen pairs had been delisted by the refetch and appear
only as absences, 6 of the remaining 179 fall short of the relaxed cut,
leaving 173 crypto pairs scored on ${\sim}11$ folds each, while only 52
of the 187 stocks clear the cut. The model-free statistic needs no
folds and covers all 179 surviving crypto pairs and all 187 stocks.

\paragraph{US-listed instruments (Alpaca)}
Fifteen-minute and one-minute bars (IEX feed), unadjusted, full 24-hour
session as distributed; RTH cells restrict to 09:30--16:00
America/New\_York, DST-aware. GBTC and ETHE traded OTC before their ETF
conversions and the IEX feed carries no OTC prints, so both begin at
uplisting (2024-01-11, 2024-07-23). The spot-Bitcoin funds all list from
2024-01-11 and the spot-Ether funds from 2024-07-23, so every one of
them covers the 2025-01--2026-02 panel span in full; the launch-era
cells of Table~\ref{tab:events} score IBIT, FBTC and GBTC over their
first six months of listed trading. BITU and ETHU obtain their
$2\times$ exposure through swaps and cash-settled futures rather than
spot custody.

\paragraph{Spot metals and FX (Dukascopy)}
Bid-side 15-minute candles with closed-market bars
(\texttt{high==low}, weekends/holidays) removed; platinum and palladium
under the feed's commodity instrument ids. The same source supplies the
bid- and ask-side series behind the quote-midprice robustness bars.

\paragraph{Transaction-cost bands}
The cost bands of Section~\ref{sec:friction} are round-trip spot
figures: ${\approx}5$\,bp at maker fees and 10--20\,bp at taker fees,
bracketing published retail and mid-tier schedules on the major venues
over the sample. Fees rather than spread are what bind
(\S\ref{sec:friction}); the wider Dukascopy quoted spreads of
Appendix~\ref{app:robust} are a retail contract-for-difference venue's
and are not used to price this friction.

\section{Reproducibility}\label{app:repro}
Every number, table and figure is produced by a pipeline of small
Python modules, distributed as the public replication package linked
in the data-availability statement (pipeline code, frozen symbol lists
and result files with SHA-256 hashes, fetch scripts rebuilding every
input from public sources, and a README mapping each exhibit to one
command); seeded procedures make reruns deterministic. The module
names quoted throughout the text (\texttt{wide}, \texttt{signlag},
\texttt{holdout\_did}, \texttt{panel\_fdr}, \texttt{depth\_test},
\ldots) are those modules; the README maps every exhibit to its
command.

\end{document}